\documentstyle[12pt]{article}
\begin{document}
\def\thefootnote{\fnsymbol{footnote}}
\begin{flushright}
KANAZAWA-98-11  \\ 
August, 1998
\end{flushright}
\vspace*{2cm}
\begin{center}
{\LARGE\bf Vacuum Structure
of the $\mu$-Problem Solvable Extra U(1) Models}\\
\vspace{1 cm}
{\Large Daijiro Suematsu}
\footnote[1]{Email: suematsu@hep.s.kanazawa-u.ac.jp}
\vspace {1cm}\\
{\it Department of Physics, Kanazawa University,\\
        Kanazawa 920-1192, Japan}\\    
\end{center}
\vspace{1cm}
{\Large\bf Abstract}\\  
Vacuum structure and related phenomenological features are 
investigated in $\mu$-problem solvable supersymmetric extra 
U(1) models.
We present a framework for the analysis of their vacuum
structure taking account of an abelian gauge kinetic term mixing,
which can potentially modify a scalar potential and 
$Z^0$ gauge interactions. 
Applying this to data of the precise measurements at LEP, 
we constrain an allowed region in a space of Higgs 
vacuum expectation values based on consistency with potential minimum 
conditions. 
We find that such a region is confined into rather restricted one.
Bounds on masses of an extra U(1) gauge boson and 
the lightest neutral Higgs boson are predicted.
Renormalization group equations for gauge coupling constants and
gaugino soft masses in an abelian gauge sector are also discussed in
relation to the gauge kinetic term mixing in some detail.
\newpage
\setcounter{footnote}{0}
\def\thefootnote{\arabic{footnote}}
\section{Introduction}
Although the standard model (SM) has been confirmed in incredible 
accuracy through precise measurements at LEP,
it is still not considered as the fundamental theory of particle
physics and its various extensions have been proposed.
One direction of such extensions is supersymmetrization of the SM from 
a viewpoint of gauge hierarchy problem.  
For this problem it is now considered
as the most promising extension of the SM \cite{rev}.
Another direction is the extension of gauge structure and it is 
represented by GUT models like SU(5) and SO(10). 
Among such extensions the simplest one is an addition of an 
extra U(1) factor group to the SM gauge structure.  
It is an interesting aspect of this extension that this kind of 
gauge structure often appears in the low energy effective models 
of perturbative superstring \cite{dflat}. 

Even in the supersymmetrized models 
a theoretically unsatisfactory feature remains from the viewpoint of 
naturalness. This is known as a $\mu$-problem \cite{mu}. 
The supersymmetric SM has a supersymmetric Higgs mixing term 
$\mu H_1H_2$. In order to induce the weak scale correctly, 
we should keep $\mu\sim O(G_F^{-1/2})$ by hand, where $G_F$ is a
Fermi constant. On the other hand,
in the supersymmetric models a typical low energy scale is generally 
characterized by a supersymmetry breaking scale $M_S$ in an
observable sector.  
There is no reason why $\mu$ should be such a scale since $\mu$
parametrizes a supersymmetric term and then it seems rather natural
to take it as a cut-off scale like $M_{\rm Pl}$.
 
A reasonable way to answer this question is to consider an origin 
of $\mu$ scale as a result of supersymmetry breaking \cite{superb}.
One of such solutions is an introduction of a
singlet field $S$ and replace $\mu H_1H_2$ by a Yukawa type coupling
$\lambda SH_1H_2$ \cite{singlet}. 
If $S$ gets a vacuum expectation value (VEV) of order
1~TeV as a result of both supersymmetry breaking and radiative 
corrections to soft supersymmetry breaking terms \cite{rad}, 
$\mu\sim O(G_F^{-1/2})$ will be dynamically realized 
through $\mu=\lambda\langle S\rangle$. 
It is noticable that the models extended by an extra U(1) symmetry
which is broken by a SM singlet field $S$ can have this feature
[7 - 12].
The existence of this U(1) can also make it free from the tadpole
problem usually unavoidable in the models with gauge singlet Higgs scalars. 
Thus the supersymmetric SM extended with an extra U(1) symmetry can be
considered as one of the simplest and most promising extensions of the 
supersymmetric SM. 
This kind of models have various interesting features and
their phenomenological aspects have been studied by various 
authors [7 - 12].

It is worthy to note that the extra U(1) models can have an another
interesting feature.
In principle, there can be a kinetic term mixing among abelian gauge 
fields because their field strength is gauge invariant.
A decade ago it was suggested that such a mixing might appear in
suitable unified models \cite{hold} and also in the effective theory 
of perturbative superstring \cite{ms,mix2}.
Following these works, in the supersymmetric models extended 
with an extra U(1), the running of gauge coupling constants \cite{run} and 
also the effects on the electroweak parameters \cite{electro} 
due to this mixing have been studied.
Recently the relation of the kinetic term mixing to the 
leptophobic property and the electroweak parameters 
has also been intensively studied in
E$_6$ inspired extra U(1) models \cite{kinet,kinet2,riz}.

In supersymmetric models gauge fields are embedded in the vector
superfields. This means that the similar effect appears also in other
component fields contained in the vector superfields, that is,  
gauginos and auxiliary fields $D$.
Recently, it has been shown that gauge kinetic term mixing can cause
additional interesting effects in various phenomena through the
neutralino sector \cite{neutralino1,neutralino2}.
However, there still seems to remain an unstudied interesting aspect 
due to the modification of $D$ fields related 
to the vacuum structure.
In this paper we investigate its effect on the scalar 
potential and the gauge interaction sector to examine the vacuum
structure of such models.
As one of its results, we can predict the bounds for 
masses of an extra neutral gauge boson and the lightest neutral 
Higgs boson.
 
The organization of this paper is as follows.
In section 2 we review general features of physical effects of
the gauge kinetic term mixing in the models with U(1)$_a \times$
U(1)$_b$ gauge symmetry from various points of view.
In section 3 we introduce the $\mu$-problem solvable extra U(1)
models studied in this paper.
In section 4 the discussion in section 2 are applied to formulate a
framework for the study of a vacuum structure of the models in the 
basis of both the results of precise measurements at LEP and the potential
minimization. Using numerical analysis based on this framework 
we constrain an allowed region in a space extended by the 
Higgs VEVs. 
We also discuss the mass bounds of the extra neutral gauge boson and
the lightest neutral Higgs boson.
Section 5 is devoted to the summary. In an appendix we review the
derivation of the electroweak parameters in the extra U(1) models.

\section{ Kinetic term mixing}
\subsection{General feature}
At first we review the general features of gauge kinetic 
term mixing in the case of U(1)$_a\times$ U(1)$_b$ model \cite{kinet}.
This is also aimed to fix various notations used in the following arguments.
The Lagrangian considered here is written as
\begin{equation}
{\cal L}=-{1\over 4}\hat F^{a\mu\nu}\hat F^a_{\mu\nu}
-{1 \over 4}\hat F^{b\mu\nu}\hat F^b_{\mu\nu}
-{\sin\chi \over 2}\hat F^{a\mu\nu} \hat F^b_{\mu\nu}
+\vert \hat D_\mu \phi\vert^2+i\bar\psi\gamma^\mu \hat D_\mu \psi.
\end{equation}
where $\sin\chi$ parametrizes the gauge kinetic term mixing.

A covariant derivative $\hat D_\mu$ is defined as $ \hat D_\mu=\partial_\mu 
-ig_a^0Q_a\hat A_\mu^a-ig_b^0Q_b\hat A_\mu^b$.
We can change this Lagrangian into the one written in a canonically 
normalized basis by resolving this
mixing in terms of the following transformation,
\begin{equation}
\left(\begin{array}{c} \hat A_\mu^a \\ \hat A_\mu^b \\ \end{array}\right)
=\left(\begin{array}{cc}1 & -\tan\chi \\ 0 & 1/\cos\chi \\ 
\end{array}\right)\left(\begin{array}{c} A_\mu^a \\ A_\mu^b \\
\end{array}\right).
\end{equation}
The resulting Lagrangian can be expressed as
\begin{equation}
{\cal L}=-{1\over 4}F^{a\mu\nu}F^a_{\mu\nu}
-{1 \over 4}F^{b\mu\nu}F^b_{\mu\nu}
+\vert D_\mu \phi\vert^2+i\bar\psi\gamma^\mu D_\mu \psi.
\end{equation}
Now the covariant derivative is altered into 
\begin{equation}
D_\mu=\partial_\mu -ig_aQ_aA_\mu^a-i\left(g_{ab}Q_a+g_bQ_b\right)A_\mu^b,
\end{equation}
where 
\begin{equation}
g_a=g_a^0,\quad g_{ab}=-g_a^0\tan\chi, \quad  g_b={g_b^0 \over
\cos\chi}.
\end{equation}
Physical phenomena should be considered by using this new
Lagrangian.
The mixing effects are confined into
the interactions between the U(1)$_b$ gauge field and matter fields.
In this new Lagrangian a gauge coupling constant $g_b$ is varied from
the original one and also a new off-diagonal gauge coupling $g_{ab}$
appears. 

Here we should note that Eq. (2) is not a unique choice of the basis
which resolves the mixing. There is an additional freedom of the
orthogonal transformation. Another useful basis $(\bar A^a_\mu, \bar A^b_\mu)$ 
is related to the basis $(A^a_\mu, A^b_\mu)$ by the orthogonal transformation.
By fixing the definition of charges $(Q_a, Q_b)$ of U(1)$_a \times$ 
U(1)$_b$, these bases are
related by the orthogonal rotation as follows, 
\begin{eqnarray}
&&\left( \begin{array}{c}
A_a \\ A_b \\ \end{array}\right)=
\left(\begin{array}{cc}\cos\theta & -\sin\theta\\
\sin\theta & \cos\theta \end{array}\right)\left( \begin{array}{c}
\bar A_a \\ \bar A_b \\ \end{array}\right), \nonumber \\ 
&&\left(\begin{array}{cc}g_a & g_{ab}\\
0 & g_b \end{array}\right)=
\left(\begin{array}{cc}\bar g_a & \bar g_{ab}\\
\bar g_{ab} & \bar g_b \end{array}\right)\left( \begin{array}{cc}
\cos\theta &\sin\theta \\ -\sin\theta & \cos\theta \\ \end{array}\right).
\end{eqnarray}
We call this new basis as the symmetric basis hereafter. Under this
basis the covariant derivative is represented as
$$
D_\mu=\partial_\mu -i\left(\bar g_aQ_a+\bar g_{ba}Q_b\right)\bar A_\mu^a
-i\left(\bar g_{ab}Q_a+\bar g_bQ_b\right)\bar A_\mu^b.
$$

These changes induced by resolving the mixing 
can bring various effects on the low energy phenomena.
A typical example of such effects is a radiative correction to gauge 
coupling constants.
For the study of the running of abelian gauge coupling constants, these
effects should be taken into account.  
In the present model the one-loop renormalization group equations (RGEs) for
these couplings 
generally take a $2\times 2$ matrix form \cite{run}.
If we use $t=\ln M/M_0$ where $M$ is a renormalization
point, these can be written as
\begin{equation}
{d\over dt}\left(\begin{array}{cc}
\bar g_a&\bar g_{ab}\\ \bar g_{ba}&\bar g_b \\ \end{array}\right)=
\left(\begin{array}{cc}
\bar g_a&\bar g_{ab}\\ \bar g_{ba}&\bar g_b \\ \end{array}\right)
\left(\begin{array}{cc}
\bar\beta_{a}&\bar\beta_{ab}\\ \bar\beta_{ab}&\bar \beta_{b} \\
\end{array}\right),
\end{equation}
in the symmetric basis.
On the other hand, using the basis taken in Eq. (4),
RGEs are expressed as \cite{kinet},
\begin{equation}
{d\over dt}\left(\begin{array}{cc}
g_a& g_{ab}\\ 0& g_b \\ \end{array}\right)=
\left(\begin{array}{cc}
g_a&g_{ab}\\ 0& g_b \\ \end{array}\right)
\left(\begin{array}{cc}
\beta_{a}&\beta_{ab}\\ 0 &\beta_{b} \\
\end{array}\right).
\end{equation}
The rotation angle $\theta$ in Eq. (6) changes with the energy 
scale as 
\begin{equation}
{d\theta \over dt}=\bar\beta_{ab}\cos 2\theta+{1\over 2} 
(\bar\beta_a -\bar\beta_b)\sin 2\theta .
\end{equation}
Although $\beta$-functions in Eqs. (7) and (8) depend on the matter contents 
in a considering model,
their general forms in the symmetric basis can 
easily written down \cite{run}.

In connection with the running of gauge coupling constants
it will also be useful to comment on the relation between the 
charge normalization and the initial condition for RGEs study.
In the usual unified models based on a simple group, 
there is no kinetic term mixing at the
unification scale $M_U$ among
its low energy abelian factor groups.
However, their kinetic term mixing can appear at the lower energy region
through the multiplets splitting due to the symmetry breaking
at the intermediate scale \cite{hold}.
In this case if we assume a
unified coupling constant to be $g_U$, the abelian charges may be
normalized at the unification scale as usual, 
\begin{equation}
g_U^2=g_a^{02}{\rm Tr}~Q_a^2=g_b^{02}{\rm Tr}~Q_b^2.
\end{equation}
On the other hand, if we consider the models derived from superstring
the kinetic term mixing can occur even at the string scale or
unification scale $M_U$ \cite{ms,mix2}. In that case
Eq. (10) should be modified as\footnote{
In the definition of $\sin\chi$ in Eq. (1) the diagonal part is
assumed to be canonically normalized.
If there are also some corrections to them, these effects should be taken
into account in the initial conditions (11) of the coupling constants
at the unification scale \cite{mix1}.}
\begin{equation}
g_U^2=g_a^{2}{\rm Tr}~Q_a^2={g_a^2g_b^2 \over g_{ab}^2+g_a^2}{\rm Tr}~Q_b^2,
\end{equation}
where we substitute the relations (5) into Eq. (10).
This equation shows that the initial values of gauge couplings
can be shifted.
Usually charges are normalized as ${\rm Tr} Q_a^2={\rm Tr} Q_b^2$
and thus $g_a^0=g_b^0$ is satisfied. However, Eq. (11) shows
$g_a<g_b$ even at $M_U$ as a result of the kinetic term mixing
due to some dynamics above the string scale.

In this model an additional mass mixing between $A_\mu^a$ and
$A_\mu^b$ also generally appears after
spontaneous symmetry breaking of U(1)$_a\times$ U(1)$_b$ due to some
VEVs of suitable Higgs scalar fields.
This mixing can be resolved by the orthogonal transformation
\begin{equation}
\left( \begin{array}{c}
A_1^\mu \\ A_2^\mu \\ \end{array}\right)=
\left(\begin{array}{cc}
\cos\xi &\sin\xi\\ -\sin\xi&\cos\xi \\ \end{array}\right)
\left(\begin{array}{c} A_a^\mu \\ A_b^\mu\\ \end{array}\right).
\end{equation}
If we write the mass matrix as
\begin{equation}
\left(\begin{array}{cc}m_a^2&m_{ab}^2 \\ m_{ab}^2&m_b^2\\
\end{array}\right),
\end{equation}
the mixing angle $\xi$ can be written as,
\begin{equation}
\tan 2\xi={-2m_{ab}^2 \over m_b^2-m_a^2}.
\end{equation}

\subsection{Supersymmetric extension}
We now consider a supersymmetric extension \cite{rev} of the discussion in 
the previous subsection.
In that case gauge fields are extended to vector superfields 
\begin{equation}
V_{\rm WZ}(x, \theta, \bar\theta)=-\theta\sigma_\mu\bar\theta V^\mu
+i\theta\theta\bar\theta\bar\lambda-i\bar\theta\bar\theta\theta\lambda
+{1\over 2}\theta\theta\bar\theta\bar\theta D,
\end{equation}
where we use the Wess-Zumino gauge.
A gauge field strength is included in the chiral superfield
constructed from $V_{\rm WZ}$ in the well known procedure,
\begin{eqnarray}
W_\alpha(x, \theta)&=&(\bar D\bar D)D_\alpha V_{\rm WZ} \nonumber \\
&=&4i\lambda_\alpha-4\theta_\alpha
D+4i\theta^\beta\sigma_{\nu\alpha\dot\beta}
\sigma^{\dot\beta}_{\mu\beta}(\partial^\mu V^\nu-\partial^\nu V^\mu)
-4\theta\theta\sigma_{\mu\alpha\dot\beta}\partial^\mu\bar\lambda^{\dot\beta}.
\end{eqnarray}
Using these superfields the supersymmetric 
Lagrangian can be written as
\begin{equation}
{\cal L}={1\over 32}\left(W^\alpha W_\alpha\right)_F
+\left(\Phi^\dagger \exp(2g_0QV_{\rm WZ})\Phi\right)_D, 
\end{equation}
where $\Phi=(\phi, \psi, F)$ is a chiral superfield which
corresponds to a matter field.
It is convenient  to present a component
representation of each term for the following arguments,
\begin{eqnarray}
&&{1\over 32}\left(W^\alpha W_\alpha\right)_F=
-{1\over 4}V_{\mu\nu}V^{\mu\nu}-{1 \over 2}i\lambda^\alpha
\sigma_{\mu\alpha\dot\gamma}\partial^\mu\bar\lambda^{\dot\gamma}
-{1 \over 2}i(\partial^\mu\bar\lambda_{\dot\beta})
\bar\sigma_{\mu}^{\dot\beta\alpha}\lambda_\alpha+{1\over 2}D^2, \\
&&\left(\Phi^\dagger \exp(2g_0QV_{\rm WZ})\Phi\right)_D=
\vert
D_\mu\phi\vert^2-i\left(\bar\psi_{\dot\beta}\bar\sigma_{\mu}^{\dot\beta\alpha}
D^\mu\psi_\alpha\right)+g_0Q\phi^*D\phi \nonumber \\
&&\hspace*{4cm}+i\sqrt 2g_0Q\left(\phi^*\lambda\psi
-\bar\lambda\bar\psi\phi\right)+\vert F\vert^2,
\end{eqnarray}
where $D_\mu$ is an original covariant derivative.

If we take account of this Lagrangian,
the introduction of the gauge kinetic term mixing is
straightforward for U(1)$_a\times$ U(1)$_b$ model.
Supersymmetric gauge kinetic terms are obtained by using chiral 
superfields $\hat W_\alpha^a$ and 
$\hat W^b_\alpha$ for U(1)$_a\times$ U(1)$_b$ as
\begin{equation}
{1\over 32}\left(\hat W^{a\alpha} \hat W^a_\alpha\right)_F
+{1\over 32}\left(\hat W^{b\alpha} \hat W^b_\alpha\right)_F
+{\sin\chi\over 16}\left(\hat W^{a\alpha} \hat W_{\alpha}^b\right)_F.
\end{equation}
These can be canonically diagonarized by using the supersymmetric
version of the transformation (2).
In the supersymmetric case this transformation affects not only the gauge
field sector as Eq. (4) but also the sector of 
gauginos $\hat\lambda_{a,b}$ and 
auxiliary fields $\hat D_{a,b}$. 
These effects can be summarized as
\begin{eqnarray}
&&g_a^0Q_a\hat \lambda^a+g_b^0Q_b\hat
\lambda^b=g_aQ_a\lambda^a+\left(g_{ab}Q_a+g_bQ_b\right)\lambda^b,
\nonumber \\
&&g_a^0Q_a\hat D^a+g_b^0Q_b\hat
D^b=g_aQ_aD^a+\left(g_{ab}Q_a+g_bQ_b\right)D^b, 
\end{eqnarray}
where $\lambda_{a,b}$ and $D_{a,b}$ are canonically normalized fields.
New gauge coupling constants are represented by Eq. (5).
From these formulae we can extract various physical results.

Equations of motion for the auxiliary fields $D_{a,b}$ can be easily
derived as
\begin{equation}
D_a=-\sum_ig_aQ_a^i\vert \phi_i\vert^2, \qquad
D_b=-\sum_i\left(g_{ab}Q_a^i+g_bQ_b^i\right)\vert\phi_i\vert^2.
\end{equation} 
The D-term contribution to the scalar potential is expressed as
$V_D={1\over 2}D_a^2+{1\over 2}D_b^2$ so that the kinetic term mixing
can clearly affect the vacuum structure of the model.
When we introduce the Fayet-Iliopoulos D-terms $\xi_a\hat D_a+\xi_b\hat
D_b$ to the original Lagrangian, the above expressions for the auxiliary
fields $D_{a,b}$ will be modified as,
\begin{equation}
D_a\rightarrow D_a-\xi_a, \qquad 
D_b\rightarrow D_b-\left(\xi_a{g_{ab}\over g_a^0}
-\xi_b{g_b\over g_b^0}\right).
\end{equation} 
If we substitute this modification into the scalar potential,
the D-term contribution to the scalar potential is obtained as
\begin{equation}
V_D={1\over 2}\tilde D_a^2+{1\over 2}\left({g_{ab}\over g_a^0}\tilde
D_a+{g_b\over g_b^0}\tilde D_b\right)^2,
\end{equation}
where $\tilde D_{a,b}$ have the same form as the ones of no kinetic
term mixing,  
\begin{equation}
\tilde D_a=\sum_ig_a^0Q_a^i\vert \phi_i\vert^2+\xi_a, \qquad
\tilde D_b=\sum_ig_b^0Q_b^i\vert\phi_i\vert^2+\xi_b.
\end{equation}
The minimum of the D-term contribution is $\tilde D_a=\tilde D_b=0$,
which equals to the one in the case of $\sin\chi=0$.
However, the minimum of the total scalar potential including an F-term
contribution is expected to be modified.
Although we donot consider anomalous U(1) models here,
this also may have an interesting effect on such models.

\subsection{RGEs of abelian gauge sector}
The RGEs for the abelian gauge sector are affected by the kinetic
term mixing like non-supersymmetric case.
Here we should add some arguments on the RGEs of gauge coupling constants and
gaugino masses.
In the supersymmetric case we can write down the concrete form of 
one-loop RGEs for gauge coupling constants in a compact form \cite{kinet},
\begin{eqnarray}
{dg_a\over dt}&=&{1\over 16\pi^2}g_a^3B_{aa}, \nonumber \\
{dg_{ab}\over dt}&=&{1\over 16\pi^2}\left(g_{ab}^3B_{aa}+2g_{ab}^2g_bB_{ab}
+g_{ab}g_b^2B_{bb}+2g_{a}^2g_{ab}B_{aa}+2g_a^2g_bB_{ab}\right),
\nonumber \\
{dg_b\over dt}&=&{1 \over 16\pi^2}\left(g_bg_{ab}^2B_{aa}
+2g_b^2g_{ab}B_{ab}+g_b^3B_{bb}\right),
\end{eqnarray}
where we use the asymmetric basis defined by Eq. (2).\footnote{
The rotation angle $\theta$ which relates both basis in 
Eq. (9) satisfies~
$\displaystyle {d\theta \over dt}={1\over 16\pi^2}
\left(g_ag_{ab}B_{aa}+g_ag_bB_{ab}\right).$}
The usage of this basis is convenient for the practical purpose
in the unified models
because U(1)$_a$ coupling is not altered from the original one
as shown in Eq. (5).
The charge factor $B_{ij}$ is defined by $B_{ij}={\rm Tr}(Q_iQ_j)$
where $Q_i$ and $Q_j$ are U(1)$_a$ or U(1)$_b$ charges of the chiral
superfields
which run through the internal line. The trace should be taken 
for all possible chiral superfields in the loop.

As mentioned before, in the ordinary unification of the abelian 
factor groups into a simple group where the usual initial condition
(10) is used, both of an initial value $g_{ab}$ and $B_{ab}$ vanish and
nonzero $g_{ab}$ can never appear at the low energy region if any
multiplet splitting is not induced by some symmetry breakings at the
intermediate scale.
However, even in such models if there appear incomplete multiplets 
of the unification group at some intermediate region, 
$B_{ab}$ becomes nonzero as its result and $g_{ab}$ will develope nonzero 
value at the lower energy region \cite{hold}.
Recently detailed analysis on the magnitude of the induced kinetic
term mixing has been done in \cite{riz} for the E$_6$ inspired extra
U(1) models.
In perturbative superstring models nonzero $g_{ab}$ can also 
appear even at the Planck scale
as pointed out in ref. \cite{ms,mix2}.
In that case we should use Eq. (11) as an initial condition for the
RGEs study.

One-loop RGEs for U(1)$_a\times$ U(1)$_b$ gaugino masses take 
also a $2\times 2$ matrix form
and can be written in the symmetric basis as,
\begin{equation}
{d\over dt}\left(\begin{array}{cc}
M_a&M_{ab}\\ M_{ab}&M_b \\ \end{array}\right)=
-\left(\begin{array}{cc}
M_a&M_{ab}\\ M_{ab}&M_b \\ \end{array}\right)
\left(\begin{array}{cc}
\gamma_m^{a}&\gamma_m^{ab}\\ \gamma_m^{ab}&\gamma_m^{b} \\ \end{array}\right),
\end{equation}
where
\begin{eqnarray}
\left(\begin{array}{cc}
\gamma_m^{a}&\gamma_m^{ab}\\ \gamma_m^{ab}&\gamma_m^{b} \\ \end{array}
\right)&=&
-2\left(\begin{array}{cc}\bar \beta_a &\bar \beta_{ab}\\
\bar \beta_{ab}&\bar \beta_b\\ \end{array}\right) \nonumber \\
&=&
-{1\over 8\pi^2}\left(\begin{array}{cc}
\bar g_a&\bar g_{ba}\\ \bar g_{ab}&\bar g_b \\ \end{array}\right)
\left(\begin{array}{cc}
B_{aa}&B_{ab}\\ B_{ab}&B_{bb} \\ \end{array}\right)
\left(\begin{array}{cc}
\bar g_a&\bar g_{ab}\\ \bar g_{ba}&\bar g_{b} \\ \end{array}\right).
\end{eqnarray}
These RGEs show that the abelian gaugino mass mixing can appear 
at the low energy region as a result of the kinetic term mixing
even if there is no mixing in the initial values of the soft supersymmetry 
breaking gaugino masses.
From Eqs. (27) and (28), we cannot generally expect the unification 
relation  which is usually predicted among the gaugino masses,
if there is the kinetic term mixing. 

In addition to these radiative effects 
the existence of an abelian off-diagonal gaugino mass may appear at the
unification scale $M_U$ in 
relation to the origin of soft supersymmetry breaking.
In the $N=1$ supergravity framework it is well known
that gaugino masses are expressed at the unification scale as \cite{sugra} 
\begin{equation}
M_a={1\over 2}(Re~f_a)^{-1}F^j\partial_jf_a,
\end{equation}
where $F^j$ is the auxiliary fields in a chiral superfield $\Phi^j$
and its VEV induces the supersymmetry breaking.
In principle, the gauge kinetic function $f_a$ can have
nonzero off-diagoinal elements $f_{ab}$ for abelian factor groups
U(1)$_a\times$ U(1)$_b$.
The existence of such off-diagonal elements $f_{ab}$ was pointed out
at the one-loop effect in the perturbative superstring \cite{mix2}.
If $f_{ab}$ has the $\Phi^j$ dependence in the case of $F^j\not=0$,
nonzero $M_{ab}$ is expected to appear at $M_U$. 
This means that there may be a mixing even in the initial
condition of the RGEs for the abelian gaugino masses.
The abelian gaugino mass mixing originated from the kinetic term mixing
may be one of the interesting aspects of soft supersymmetry breakings.
This point seems not to have been noted by now.
Although further study of these issues seem to be worthy to clarify
the detailed feature of extra U(1) models, they are beyond our present 
scope and we will not treat them in this paper.

\section{$\mu$-problem solvable models}
There can be a lot of low energy extra U(1)$_X$ models
as the extension of the MSSM.\footnote{Hereafter we will use the same
notation for the chiral superfields and its scalar component fields.}
Among these models we are especially interested in $\mu$-problem solvable
extra U(1)$_X$ models, which satisfy the features such as,
(i) the extra U(1)$_X$ symmetry should be broken by the VEV of a SM
singlet scalar $S$ and (ii) the singlet chiral superfield $S$ 
has a coupling to the ordinary 
Higgs doublet chiral superfields $H_1$ and $H_2$ such as $\lambda SH_1H_2$.
In these models the ordinary $\mu$ term is forbidden in the original
Lagrangian by U(1)$_X$ and the $\mu$ scale is naturally 
related to the mass of the extra U(1)$_X$ gauge boson. 
Thus they also seem to be very interesting
from the phenomenological viewpoint.\footnote{
There is also a possibility that the $\mu$ term is induced by a
nonrenormalizable term $\lambda (S\bar S/M_{\rm pl}^2)^nSH_1H_2$
in the superpotential 
because of some discreate symmetry \cite{discr}. In such a case 
$\langle S\rangle$ should be large in order to realize the appropriate 
$\mu$ scale. As a result there is no low energy 
extra gauge symmetry which can be relevant to the present experimental 
front. The tadplole problem accompanying this singlet scalar $S$
may again appear. 
Because of these reasons we do not consider this possibility here.} 
Thus it will be worthy to prepare the framework for their 
analysis and to investigate detail features of such typical models.

In this paper we confine our attention to this class of models derived 
from the superstring inspired $E_6$ models.\footnote{It is well known that
this type of model often appears in models derived from the various
construction of weak coupling superstring \cite{dflat,string}.
The following discussion can be generalized to such models straightforwardly.} 
There are two classes of extra U(1)$_X$ models derived from superstring
inspired $E_6$ models.
One is a rank five model, which is called as $\eta$ model.
The other ones have a rank six and there are two extra U(1)s 
in addition to the SM gauge structure.
They are known to be expressed as suitable linear combinations of
two abelian groups U(1)$_\psi$ and U(1)$_\chi$.  
Their charge assignments for ${\bf 27}$ of $E_6$ are summarized in Table 1.
\begin{figure}[bt]
\begin{center}
\begin{tabular}{|c|c|c||c|c||c|c|}\hline
fields &$SU(3)\times SU(2)$& $Y$ & $Q_\psi$ & $Q_\chi$ & $Q_\eta$ 
& $Q_{\xi_\pm}$  \\
\hline
$Q$  &(3,2)&${1\over 3}$&$\sqrt{5\over 18}$&
$-{1\over\sqrt6}$&$-{2\over 3}$ &$\pm {1\over\sqrt6}$  \\    
$\bar U$&$(3^\ast,1)$&$-{4\over 3}$&$\sqrt{5\over 18}$&$-{1\over\sqrt
6}$&$-{2\over 3}$&$\pm{1\over\sqrt6}$ \\
$\bar D$&$(3^\ast,1)$&${2\over 3}$&$\sqrt{5\over 18}$&${3\over\sqrt 6}$&
${1\over 3}$ &$\pm{2\over\sqrt 6}$\\
$L$  &(1,2)&$-1$&$\sqrt{5\over 18}$&${3\over\sqrt 6}$&${1\over 3}$ 
&$\pm {2\over\sqrt 6}$\\
$\bar E$&(1,1)&2&$\sqrt{5\over 18}$&$-{1\over\sqrt 6}$&$-{2\over 3}$
&$\pm {1\over\sqrt 6}$ \\
$H_1$&(1,2)&$-1$&$-2\sqrt{5\over 18}$&$-{2\over\sqrt 6}$&${1\over 3}$
& $\mp {3\over\sqrt 6}$ \\
$H_2$&(1,2)&1&$-2\sqrt{5\over 18}$&${2\over\sqrt 6}$&${4\over 3}$
&$\mp{2\over\sqrt 6}$ \\
$g$&(3,1)&$-{2\over 3}$&$-2\sqrt{5\over 18}$&${2\over\sqrt
6}$&${4\over 3}$&$\mp {2\over\sqrt 6}$ \\
$\bar g$&$(3^\ast,1)$&${2\over 3}$&$-2\sqrt{5\over 18}$&
$-{2\over\sqrt 6}$&${1\over 3}$&$\mp{3\over\sqrt 6}$ \\
$S$ &(1,1)&0&$4\sqrt{5\over 18}$&0&$-{5\over 3}$&$\pm {5\over\sqrt 6}$ \\
$N$ &(1,1)&0&$\sqrt{5\over 18}$&$-{5\over\sqrt 6}$&$-{5\over 3}$&
0 \\\hline
\end{tabular}
\vspace*{.2cm}
\end{center}
{\small {\bf Table 1}\hspace{.5cm}
The charge assignment of extra U(1)s which are derived from $E_6$.
These charges are normalized as $\displaystyle\sum_{i\in{\bf 27}}Q_i^2=20$.}
\vspace{.5cm}
\end{figure}
As seen from this table, there is a SM singlet $S$ which has
a coupling $\lambda SH_1H_2$.
The $\eta$-model clearly satisfies the above mentioned conditions (i)
and (ii).
On the other hand, in the rank six models thse conditions impose rather 
severe constraint on the extra U(1)$_X$ remaining at the low energy region.

In this type of rank six models a right-handed sneutrino $N^c$ also has 
to get a VEV to break the gauge symmetry into the one of the SM.
If we try to explain the smallness of the neutrino mass in this
context, $N^c$ should
get the sufficiently large VEV. To make this possible $N$ needs to have 
a massless conjugate partner $\bar N$ which is a chiral superfield 
belonging to {\bf 27}$^*$ of $E_6$.\footnote{If we impose the gauge
coupling unification on our models, we need to include other conjugate 
pairs besides $N+\bar N$ from ${\bf 27}+{\bf 27}^*$ as pointed out in
\cite{wl,riz}.}
Fortunately, it is well known that this can happen in the
perturbative string models \cite{dflat}.
In such a case, as easily seen, a sector of $(N, \bar N)$ in the 
fields space has a D-flat 
direction $\langle N\rangle=\langle \bar N\rangle$ 
and then they can get a large VEV without breaking 
supersymmetry \cite{dflat}.
This VEV $\langle N\rangle$ can induce the large right-handed Majorana 
neutrino mass through the nonrenormalizable term 
$(N\bar N)^n/M_{\rm pl}^{2n-3}$ in the superpotential
and then the seesaw mechanism is applicable to yield the small
neutrino mass\footnote{
The small Majorana neutrino mass can also be directly induced through
loop effects and/or nonrenormalizable couplings by using this 
intermediate scale $\langle N\rangle$ \cite{sue0,lang2}. 
In that case some kind of discrete symmetry should play 
an important role. Here we should also note that all the right-handed 
neutrinos need not to be heavy. 
Light sterile neutrinos is possible in this type 
of models.} as suggested in \cite{extra1,sue0}.
This D-flat direction may also be related to the inflation of universe and the
baryogenesis as discussed in \cite{infl}.
However, this introduction of $\langle N\rangle$  
usually breaks the direct relation between 
the $\mu$ scale and the mass of the extra U(1)$_X$ gauge boson
because the VEV of $N$ also generally contributes to the latter.
In order to escape this situation 
we need to select a U(1)$_X$ by taking a suitable linear 
combination of U(1)$_\psi$ and U(1)$_\chi$ to make $N$ have zero
charge of this U(1)$_X$ \cite{extra1,ms,sue0}.
This type of model is also shown in Table 1. 
The difference between $\xi_{\pm}$ 
is the overall sign and they can be identified by the redefinition of
$g_b^0$ and $\sin\chi$. In the following part we adopt the
$\xi_-$ convention.\footnote{As discussed in refs. \cite{extra1,sue0}, 
$Q_{\xi_-}$ can
also be obtained only by changing the field assignments for $Q_\chi$.
This insight allows us to construct new models, which can induce an
interesting neutrino mass matrix \cite{neut} by using the
charge assignments $Q_\chi$ and $Q_{\xi_-}$ for the
different generations \cite{sue0}. However, in this paper we shall not
consider such models for simplicity.}
Using the D-flat direction $\langle N\rangle=\langle \bar N\rangle$ 
of another extra U(1) orthogonal to this U(1)$_{\xi_-}$, 
the right-handed sneutrino gets the large VEV which breaks this extra
U(1) symmetry and induces the
large Majorana masses for the right-handed neutrinos as mentioned
above.
As a result of this symmetry breaking at the intermediate scale,
only one extra U(1)$_{\xi_-}$ remains as the low energy symmetry.

Apart from the property of the low energy extra U(1)$_X$,
whether an intermediate scale can exist or not is
a special feature which discreminate between the rank six 
$\xi_-$ model and the rank five $\eta$ model.
In the following study we will concentrate our study on low energy
features of two U(1)$_X$ models $(X=\eta,~\xi_-)$.

\section{Vacuum structure of extra $U(1)_X$ models}
\subsection{General framework}
In this section we investigate the vacuum structure of
$\mu$-problem solvable $\eta$ and $\xi_-$ models.
The electroweak gauge structure of these models is 
SU(2)$_L\times$U(1)$_Y\times$U(1)$_X$ at the low energy region.
The arguments in the previous section are straightforwardly applicable 
if we identify
$A_a^\mu$ and $A_b^\mu$ with the gauge fields $B^\mu$ of U(1)$_Y$
and $X^\mu$ of U(1)$_X$, respectively.
After resolving the kinetic term mixing of these abelian factor groups,
the canonically normalized Lagrangian of the present models is 
expressed by using the component fields as,
\begin{eqnarray}
{\cal L}&=&-{1\over 4}W_{\mu\nu}^iW^{i\mu\nu}-{1\over 4}B_{\mu\nu}
B^{\mu\nu} -{1\over 4}X_{\mu\nu}X^{\mu\nu} \nonumber \\
&&-i\lambda_W^{i\alpha}
\sigma_{\mu\alpha\dot\beta}\partial^\mu\bar\lambda_W^{i\dot\beta}
-i\lambda_B^{\alpha}
\sigma_{\mu\alpha\dot\beta}\partial^\mu\bar\lambda_B^{\dot\beta}
-i\lambda_X^{\alpha}
\sigma_{\mu\alpha\dot\beta}\partial^\mu\bar\lambda_X^{\dot\beta}
\nonumber \\
&&+{1\over 2}\sum_iD_{W^i}^2+{1\over 2}D_B^2+{1\over 2}D_X^2 \nonumber \\
&&+\vert \left(\partial_\mu -G(V_\mu)\right) \phi~\vert^2
+i\bar\psi^{\dot\beta}\sigma_{\mu\alpha\dot\beta}\left(\partial^\mu 
-G(V^\mu)\right)\psi^\alpha
+{1\over 2}\phi^*G(D)\phi \nonumber \\
&&+i\sqrt 2\left( \phi^*G(\lambda)\psi-\phi G(\bar\lambda)\bar\psi\right)
+\vert F \vert^2 + \big[ W(\Phi)+ h.c. \big]_F,
\end{eqnarray}
where $W_{\mu\nu}^i, B_{\mu\nu}$ and $X_{\mu\nu}$ are the field strengths
of SU(2)$_L$, U(1)$_Y$ and U(1)$_X$, respectively.
The chiral superfields $\Phi=(\phi,\psi,F)$ should be understood to 
represent all 
necessary matter and Higgs fields, although indices for all
quantum numbers are abbreviated.
For the convenience we use the notation for the vector suprerfields 
\begin{equation}
G({\cal F})={i\over 2}\left(g_W\tau^i{\cal F}^i_W +g_YY{\cal F}_B
+(g_{YX}Y+g_XQ_X){\cal F}_X\right),
\end{equation}
where ${\cal F}$ represents the component fields $V_\mu$, $\lambda$
and $D$ of the vector superfield $V_{\rm ZW}$ for the gauge 
group SU(2)$_L\times$ U(1)$_Y\times$ U(1)$_X$.

In the $\mu$-problem solvable models the gauge symmetry
breaking required around the weak scale is
SU(2)$_L\times$U(1)$_Y\times$U(1)$_X\rightarrow$U(1)$_{\rm em}$,
which is realized by the VEVs of Higgs scalar fields
\begin{equation}
\langle H_1\rangle=\left(\begin{array}{c}v_1\\0\end{array}\right),
\quad \langle H_2
\rangle=\left(\begin{array}{c}0\\v_2\end{array}\right),
\quad \langle S\rangle=u, 
\end{equation}
where their quantum numbers for 
SU(2)$_L\times$ U(1)$_Y\times$ U(1)$_X$ are
\begin{equation}
H_1~( \underline{2}, -1, Q_1 ),\quad  H_2~( \underline{2}, 
1, Q_2 ),\quad S~( \underline{1}, 0, Q_S ).
\end{equation}
For a correct electroweak vacuum we need to impose
$v_1^2+v_2^2=(174~{\rm GeV})^2(\equiv v^2)$.
All VEVs are assumed to be real.
The vacuum of these models is described as a point in a space
of two dimensionless parameters, $\tan\beta=v_2/v_1$ and $u/v$.
In the rest of this section we give a general framework to constrain
the allowed region in this space in terms of both electroweak precision 
measurements and scalar potential minimum conditions.
After that we use it to analyze the vacuum structure of our models.

In order to investigate the vacuum structure in the basis of the
recent experimental results
we need to determine the physical states at and below the
weak scale \cite{kinet}. 
The mass mixing between two U(1) factor groups appears 
when the spontaneous symmetry breaking occurs 
at the neighborhood of the weak scale.
In the present models the charged gauge sector is the same as that 
of the MSSM. In the neutral gauge sector we introduce 
the Weinberg angle $\theta_W$ in a usual
way,\footnote{
In the following we use the abbreviated notation
$s_W\equiv \sin\theta_W$ and $c_W\equiv \cos\theta_W$.}
\begin{equation}
Z_\mu=\cos\theta_WW_\mu^3-\sin\theta_WB_\mu, \quad
A_\mu=\sin\theta_WW_\mu^3+\cos\theta_WB_\mu.
\end{equation}
Here we use the canonically normalized basis $(Z_\mu, X_\mu)$ so
that $A_\mu$ is decoupled from the $(Z_\mu, X_\mu)$ sector.
After the spontaneous symmetry breaking the mass matrix 
of $(Z_\mu, X_\mu)$ can be written as
\begin{equation}
\left( \begin{array}{cc}
m_Y^2 & m^2_{YX} \\
m^2_{YX} & m^2_X \\
\end{array} \right),
\end{equation}
where each element is expressed as
\begin{eqnarray}
&&m_Y^2=m_Z^2, \nonumber \\
&&m_{YX}^2=m_Z^2s_W\tan\chi+{\Delta m^2 \over \cos\chi}, \nonumber \\
&&m_X^2=m_Z^2s^2_W\tan^2\chi 
+2\Delta m^2s_W{\sin\chi \over\cos^2\chi}
+{M_{Z^\prime}^2\over \cos^2\chi}.
\end{eqnarray}
In these expressions $m_Z^2, \Delta m^2$ and $M_{Z^\prime}^2$ represent 
the values of corresponding components in the case of no 
kinetic term mixing $(\sin\chi=0)$. They can be written as
\begin{eqnarray}
&&m_Z^2={1\over 2}(g^2_W+g_Y^2)v^2, \nonumber \\
&&\Delta m^2={1 \over 2}(g_W^2+g_Y^2)^{1/2}g_X
\left(Q_1v_1^2 -Q_2v_2^2\right), \nonumber \\
&&M_{Z^\prime}^2={1\over 2}g_X^2\left(Q_1^2v_1^2+Q_2^2v_2^2+Q_S^2
u^2\right).
\end{eqnarray}

We introduce mass eigenstates which diagonalize the mass matrix
(35) as follows,
\begin{eqnarray}
&&Z^\mu_1=\cos\xi Z^\mu +\sin\xi X^\mu, \nonumber\\
&&Z^\mu_2=-\sin\xi Z^\mu +\cos\xi X^\mu,
\end{eqnarray}
where the mixing angle $\xi$ can be given by using Eq. (13) as
\begin{equation}
\tan2\xi={-2\cos\chi(m_Z^2s_W\sin\chi+\Delta m^2) \over
M_{Z^\prime}^2+2\Delta m^2s_W\sin\chi
+m_Z^2s^2_W\sin^2\chi-m_Z^2\cos^2\chi}.
\end{equation}
In general, a mass eigenvalue $m_{Z_2}$ and $\xi$ are severely 
constrained by direct searches at Fermilab Tevatron \cite{tev} and
precise measurements at LEP \cite{lep}.
In the case of $\sin\chi=0$, it is usually assumed that this constraint 
is satisfied because of $\Delta m^2 \ll M_{Z^\prime}^2$.
This can be realized in two ways. 
Simple one requires only $v_1^2, v_2^2 \ll u^2$ which needs no fine
tuning at this stage \cite{extra1,cdeem}.
The other one requires a special situation $\Delta m^2\sim 0$
\cite{cdeem,kinet}.
This reduces to $\tan^2\beta\sim 1/4$ ($\eta$ model) and 3/2
($\xi_-$ model). Here it may be useful to remember that the radiative
symmetry breaking scenario favors $\tan\beta >1$.
If $\sin\chi\not=0$, however, there may be a new possibility to
realize the small $\xi$ even if $\Delta m^2\ll M_{Z^\prime}^2$
is not satisfied. Such a situation can be expected 
to occur without any assumption for the largeness of $u$ or $\Delta
m^2\sim 0$,
if the following condition is satisfied\footnote{This role of
$\sin\chi\not= 0$ can be played by the VEVs of new Higgs doublet
scalars which have nonzero charges of U(1)$_X$ \cite{kinet}. }
\begin{equation}
\sin\chi\sim -{\Delta m^2\over m_Z^2s_W}= 
-{g_X\over g_Y}{Q_1v_1^2-Q_2v_2^2 \over v^2}.
\end{equation} 
It may be useful to note that this condition is reduced to
\begin{equation}
\sin\chi\sim \left\{
\begin{array}{ll}
{\displaystyle {g_X\over 3g_Y}{4\tan^2\beta -1\over \tan^2\beta+1}}  
&\qquad \eta~{\rm model}, \\
{\displaystyle {g_X\over \sqrt 6g_Y}{2\tan^2\beta -3\over \tan^2\beta+1}}&
\qquad \xi_-~{\rm model}. \\
\end{array} \right.
\end{equation}
This shows that there is a special $\sin\chi$ value for each
$\tan\beta$ to realize $\xi=0$.
In the case of $\eta$ model $\tan\beta >1$ 
can requires $\sin\chi~{^>_\sim}~0.5g_X/g_Y$ to satisfy this
condition.\footnote{In the $\eta$ model it is known that there is a
special $\sin\chi$ value which can make U(1)$_\eta$ leptophobic
$(~\sin\chi=g_X/3g_Y~)$ \cite{kinet}. 
It corresponds to $\sin\chi=1/3$ if we take $g_X/g_Y=1$
at the weak scale.}

In the case of $M_{Z^\prime}^2 \gg m_Z^2,~ \Delta m^2$,~ 
the mass eigenvalues of Eq. (35) are written as
\begin{eqnarray}
&&m_{Z_1}^2 \simeq m_Z^2
-{1\over M_{Z^\prime}^2}\left(m_Z^2s_W\sin\chi
+\Delta m^2\right)^2 \\ 
&&m_{Z_2}^2 \simeq {M_{Z^\prime}^2\over \cos^2\chi}
+{1\over M_{Z^\prime}^2}\left(m_Z^2s_W\sin\chi
+\Delta m^2\right)^2.
\end{eqnarray}
Original states which are not canonically normalized
can be related to the mass eigenstates $({\cal A}^\mu, Z_1^\mu, Z_2^\mu)$ as
\begin{eqnarray}
&&\hat A^\mu={\cal A}^\mu -c_W\tan\chi\left(\sin\xi Z_1^\mu+\cos\xi
Z_2^\mu\right), \nonumber \\
&&\hat Z_\mu=\left(\cos\xi+s_W\tan\chi\sin\xi\right)Z_1^\mu
+\left(-\sin\xi+s_W\tan\chi\cos\xi\right)Z_2^\mu, \nonumber \\
&&\hat X^\mu={\sin\xi\over\cos\chi} Z_1^\mu 
+{\cos\xi\over \cos\chi} Z_2^\mu,
\end{eqnarray}
where ${\cal A}^\mu$ stands for a real photon field and $Z_1^\mu$ is
understood as $Z^{0\mu}$ observed at LEP.
Using these mass eigenstates, the interaction terms of ${\cal A}^\mu$
and $Z_1^\mu$ with ordinary quarks and leptons can be expressed as
\cite{kinet},
\begin{eqnarray}
&&{\cal L}_{\rm int}=eQ_{\rm em}\bar\psi\gamma_\mu\psi{\cal A}^\mu
+{g_W\over 2c_W}\left[(\bar v_f + \bar v_f^\prime\bar\xi)
\bar\psi\gamma_\mu\psi +(\bar a_f + \bar a_f^\prime\bar\xi)
\bar\psi\gamma_\mu\gamma_5\psi\right] Z_1^\mu,   \\
&&\hspace*{1cm}\bar v_f=\left({\tau^3\over 2}-2Q_{\rm em}s^2_W \right)
(1+s_W\xi\tan\chi)-2Q_{\rm em}c_W^2s_W\xi\tan\chi, \nonumber \\
&&\hspace*{1cm}\bar v_f^\prime =(Q_X^{\psi_L}+Q_X^{\psi_R}), 
\nonumber \\
&&\hspace*{1cm}\bar a_f=-{\tau^3 \over 2}(1 +s_W\xi\tan\chi), 
\nonumber \\
&&\hspace*{1cm}\bar a_f^\prime =(Q_X^{\psi_R}-Q_X^{\psi_L}),
\end{eqnarray}
where we assume $\xi$ is small enough and put $\displaystyle
\bar\xi\equiv {g_Xc_W \over g_W\cos\chi}\xi$. 
$Q_X^{\psi_L}$ and $Q_X^{\psi_R}$ stand for the U(1)$_X$ charges of
$\psi_L$ and $\psi_R$.
Using these effective couplings,
the partial decay width $\Gamma_f$ and the asymmetry
parameter $A_f$ for $Z \rightarrow \bar ff$ are expressed as
\begin{eqnarray}
&&\Gamma_f={G_F m_{Z_1}^3\over 6\sqrt 2\pi}\rho N_c\left[
(\bar v_f + \bar v_f^\prime\bar\xi)^2 +(\bar a_f + \bar
a_f^\prime\bar\xi)^2 \right], 
\nonumber \\
&&A_f={2(\bar v_f + \bar v_f^\prime\bar\xi)
(\bar a_f + \bar a_f^\prime\bar\xi)\over 
(\bar v_f + \bar v_f^\prime\bar\xi)^2+(\bar a_f + \bar a_f^\prime\bar\xi)^2},
\end{eqnarray}
where $\rho$ is a one-loop corrected ratio of the neutral current 
to the charged current and $N_c$ stands for the effective color factor.
The forward-backward asymmetry of $Z \rightarrow \bar ff$ is given as 
$\displaystyle A_{\rm FB}^f={3\over 4}A_eA_f$. 
Using these formulae, we can restrict the allowed region in a
$(\tan\beta, ~u/v)$ plane by using the electroweak data.

\subsection{Constraints from electroweak data}
The deviations of various electroweak observables in the present
models from the SM predictions 
can be strictly constrained by taking account of data obtained at LEP.
Electroweak parameters corrected by the extra U(1)$_X$ effects from the SM
values are the $\rho$ parameter and the effective Weinberg angle $\bar s_W^2$.
In these parameters we only take account of the correction induced 
by the extra U(1)$_X$ effect besides the radiative corrections in the
SM. 
Other corrections yielded by the exotic 
matter fields are model dependent and we assume that they 
are small enough \cite{kinet}.
More concretely, we put $\rho=1+\delta\rho_t +\delta \rho_M$ where
$\delta\rho_t$ is a one-loop correction due to top quark in the SM 
and $\delta\rho_M$ is the extra U(1)$_X$ effect.
The detailed derivation of both deviations $\delta\rho_M$ and
$\Delta\bar s_W^2$ due to the existence of
extra U(1)$_X$ is reviewed in the appendix \cite{electro,kinet}. 
We only present the results below.

$\delta\rho_M$ appears at the tree level as a result of both of 
the mass and kinetic term mixing. 
It can be expressed as \cite{kinet}
\begin{equation}
\delta\rho_M\simeq {M_{Z^\prime}^2 \over m_Z^2}{\xi^2\over \cos^2\chi}
+2\bar s_W\xi\tan\chi, 
\end{equation}
where we use a relation $m_Z^2/m_{Z_1}^2 \simeq 
M_{Z^\prime}^2\xi^2/m_Z^2\cos^2\chi$ which is obtained from
Eqs. (39) and (42) as far as $\xi \ll 1$.
The second term in the right-hand side comes 
from the kinetic term mixing $(\sin\chi\not=0)$.
This deviation in the $\rho$ parameter can also make an influence in
the deviation of $\bar s_W^2$.
The expression of $\Delta\bar s_W^2$ including such an effect
can be found as \cite{kinet} 
\begin{equation}
\Delta \bar s_W^2 \simeq -\bar c_W^2\xi\left({\bar s_W^2
\over \bar c_W^2-\bar s_W^2}{M_{M^\prime}^2\over m_Z^2}{\xi\over
\cos^2\chi}-\bar s_W\tan\chi\right).
\end{equation}
The experimental bounds obtained at LEP for these parameters can put 
strong constraints on the parameters related to the extra 
U(1)$_X$ sector.

Following the procedure used in Ref. \cite{prec}, we estimate the
deviation of LEP observables ${\cal O}$ from the predictions 
in the present models.
In the present analysis we use $\Gamma_Z, ~R_\ell, ~\sigma_{\rm had},
~R_b, ~R_c, ~m_W/m_Z, ~A_{\rm FB}^b, ~A_{\rm FB}^c$ and $A_{\rm FB}^\ell$ 
as ${\cal O}$.
These electroweak observables can be calculated by using the tree
level formulae (47) and
the deviation of these observables ${\cal O}$ can be approximately expanded
by $\delta\rho_M$, $\Delta \bar s_W^2$ and $\bar\xi$ as
\begin{equation}
{\delta {\cal O}\over {\cal O}}={\cal A}^{(1)}\delta\rho_M 
+{\cal A}^{(2)}\Delta\bar s_W^2 +{\cal B}\bar\xi.
\end{equation}
where ${\cal A}^{(1)}, ~{\cal A}^{(2)}$ and ${\cal B}$ are calculated
by using Eqs. (46) and (47).
As is easily checked, only ${\cal B}$ is different between both models.
Their numerical values of the present models are given in Table 2. 

\begin{figure}[tb]
\begin{center}
\begin{tabular}{|c||c|c|c|c|c|c|c|c|c|}\hline
${\cal O}$&$\Gamma_Z$& $R_{\ell}$ & $\sigma_{\rm had}$ & $R_b$ &
$R_c$ & $m_W/m_Z$ & $A^b_{\rm FB}$& $A^c_{\rm FB}$& $A^\ell_{\rm FB}$
\\\hline\hline
${\cal A}^{(1)}$&0.99&0&0&0&0&0.38&0&0&0\\ 
${\cal A}^{(2)}$&$-1.06$&$-0.84$&0.097&0.18&$-0.35$&$-1.00$&$-54.15$&
$-58.66$&$-106.93$\\
${\cal B (\eta)}$&0.52&$-0.82$&0.20&2.56&$-4.86$&0&$-26.61$&$-25.41$
&$-54.78$\\ 
${\cal B (\xi_-)}$&$0.33$&$-4.31$&$5.03$&$1.86$
&$-3.54$&0&$7.21$&$9.71$&$16.98$\\\hline
\end{tabular}
\vspace*{0.3cm}\\
{\bf Table 2}
\end{center}
\end{figure}

\normalsize
Since $\delta\rho_M$ and $\Delta\bar s_W^2$ are not independent and 
are related through Eqs. (48) and (49),
four independent free parameters $g_X,~\sin\chi,~\delta\rho_M$ and
$\bar\xi$ are contained in our analysis.
If we take $g_X$ and $\sin\chi$ to be suitable values,
we can carry out two parameters $\chi^2$-fit between these 
predictions and LEP data in a $(\delta\rho_M, ~\bar\xi)$ plane. 
As a result of such analyses a minimum value of $\chi^2$ 
is found to be $\chi^2\simeq 7.0$ for 7 degrees of freedom
at a point such as 
\begin{eqnarray}
(\delta\rho_M, \bar\xi) &\sim& (-2.4\times 10^{-4}, -4\times 10^{-5}),
\qquad \eta~{\rm model}, \nonumber \\
      &\sim& (-2.4\times 10^{-4}, ~4\times 10^{-5}),
\qquad \xi_-~{\rm model}  
\end{eqnarray}
in the case of $\sin\chi=0$.
For this degrees of freedom the corresponding $\chi^2$ values to
goodness-of-fits of 95\% and 99\% are $\chi^2=14.1$ and 18.5, respectively.
For the same observables the $\chi^2$ value of the SM is 3.4.
Thus at this stage there is no positive experimantal signature for 
considering the low energy extra U(1)$_X$ models.
These data should be only used to put the constraint on the 
vacuum structure of the models so as to satisfy the constraint on $\bar\xi$.
In the following study our aim is to find what kind of vacua 
can approximately satisfy this constraint.

By using the relations (37), (39), (48) and (49), we can project this result 
in the $(\delta\rho_M, ~\bar\xi)$ plane onto the ones in the various planes
defined by other variables.
To see a role of the kinetic term mixing on the value of $u$, 
it is convenient to draw $\chi^2$-contours
for typical values of $\tan\beta$ and $g_X$ in a $(\sin\chi, ~u/v)$ plane.
It should be reminded that the value of $\sin\chi$ which can induces
the minimum of $u$ is largely dependent on $\tan\beta$ as shown in
Eq. (41). Here we choose $\tan\beta=1.5$ taking account of the fact that 
the radiative symmetry breaking favors $\tan\beta~{^>_\sim}~1$.
These results for $\eta$ and $\xi_-$ models are presented in Fig. 1.
From this figure we can easily find that it shows very similar features 
to a contour of the present upper bound of mixing angle $\xi$ given in
\cite{neutralino2} as expected. 
In the $\eta$ model $\sin\chi=1/3$ which corresponds to 
leptophobia in the present setting cannot make $u$ small because $\xi$ 
is large for $\tan\beta=1.5$.
For this $\tan\beta$ we need $\sin\chi\sim 0.82$ to realize $\xi\sim 0$.
Here we should note that too large
$\sin\chi$ value seems to be difficult to be realized \cite{riz} and also
it may contradict a perturbative picture.

\input epsf
\begin{figure}[htb]
\begin{center}
\epsfysize=7cm
\leavevmode
\epsfbox{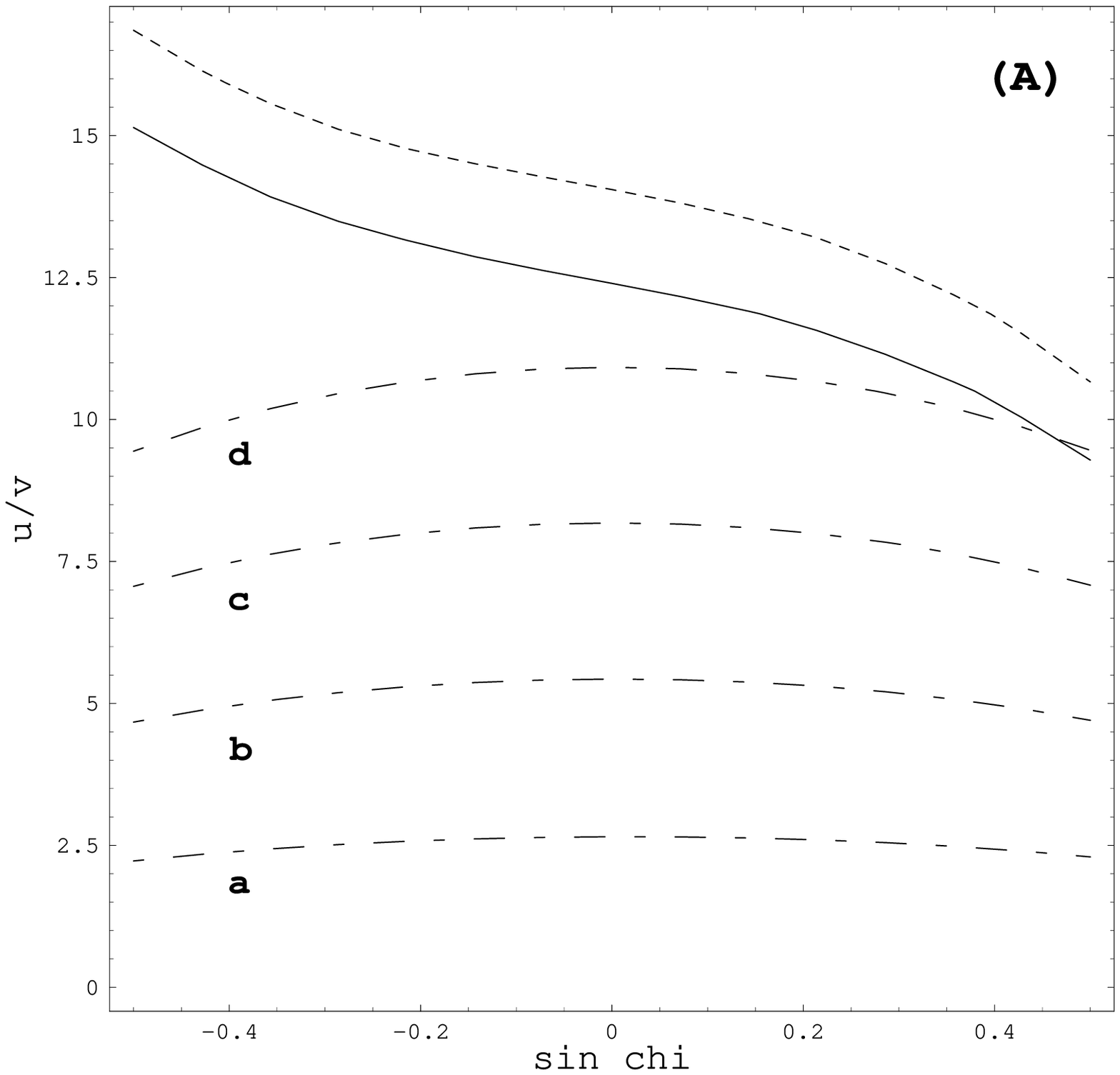}
\hspace*{1cm}
\epsfysize=7cm
\leavevmode
\epsfbox{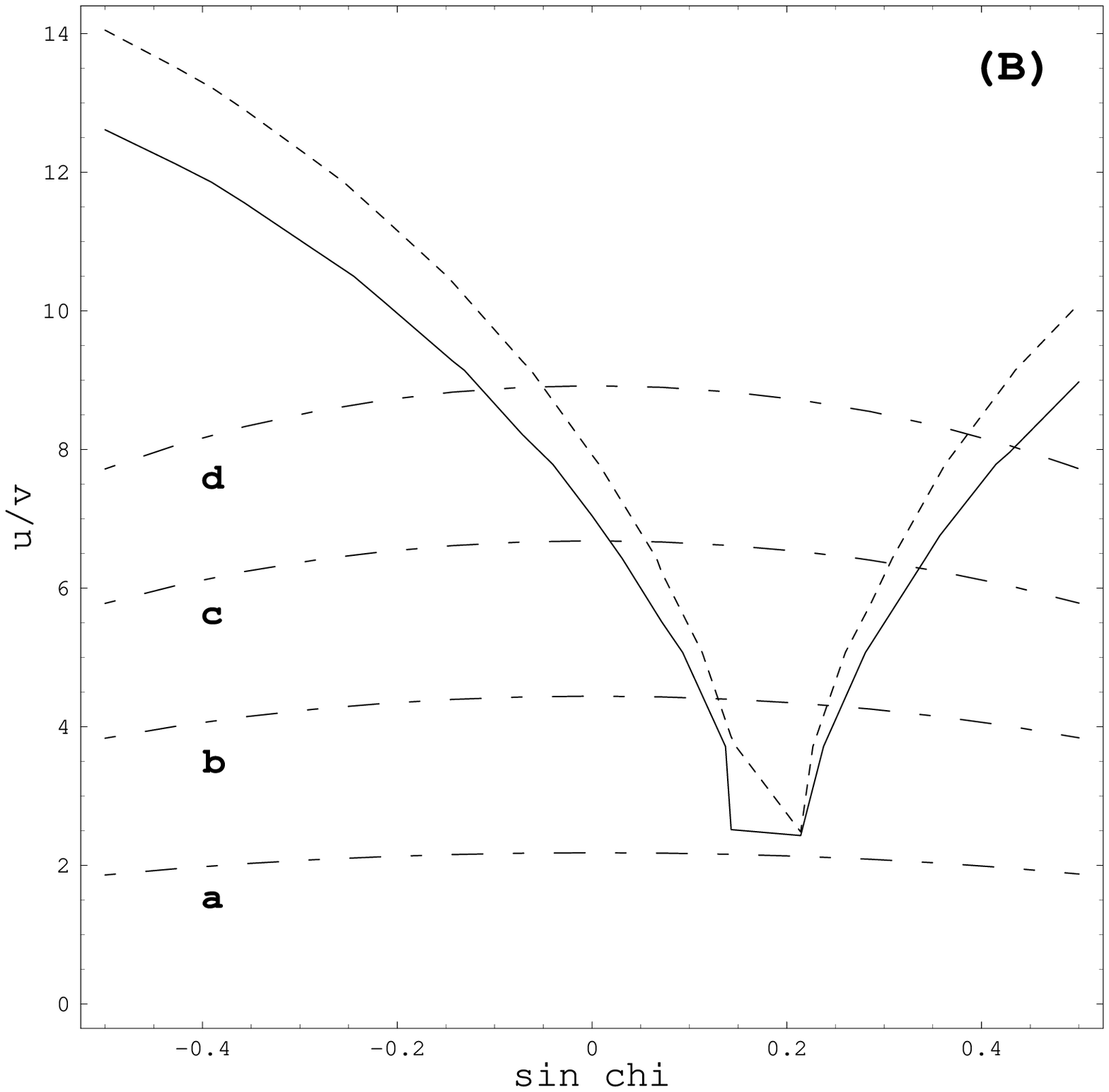}
\end{center}
\vspace*{-3mm}
{\footnotesize Fig.1\ $\chi^2$-contours in the $(\sin\chi, ~u/v)$ plane
for (A) $\eta$ and (B) $\xi_-$ models with $\tan\beta=1.5$ and $g_X/g_Y=1$.
They are shown by a solid line for $\chi^2=18.5$ and a dotted
line for $\chi^2=14.1$, respectively. 
Dash-dotted lines represent the $m_{Z_2}$ contours such as a: 400
GeV, b: 800 GeV, c: 1200 GeV and d: 1600 GeV, respectively.} 
\end{figure}

In order to examine the vacuum structure we should draw 
the $\chi^2$-contours in the $(\tan\beta, ~u/v)$ plane
since the vacuum is parameterized by two dimensionless variables
$\tan\beta$ and $u/v$.
In this analysis relevant free parameters in the model are $g_X$ and
$\sin\chi$ alone. 
We vary these parameters in the following regions,
\begin{equation}
0\le\sin\chi\le 0.3, \qquad 0.8\le g_X/g_Y\le1.2. 
\end{equation} 
The study of RGEs in Eq. (26) for various E$_6$ inspired extra U(1) models
\cite{riz} seems to verify this assumption on $\sin\chi$ and $g_X/g_Y$ at the
weak scale.
The reason to take $\sin\chi$ positive has already discussed in the
previous subsection. 
In Fig. 2 we present the $\chi^2$-contours in 
this plane for both models with typical values of these parameters.
In the same plane we also plot contours of the extra neutral gauge boson mass 
$m_{Z_2}$ as a reference.
We can observe that $u$ can be small and result in a rather small
$m_{Z_2}$ value for a suitable region of $\tan\beta$.
This $\tan\beta$ region is somehow different between two models.

\begin{figure}[htb]
\begin{center}
\epsfysize=7cm
\leavevmode
\epsfbox{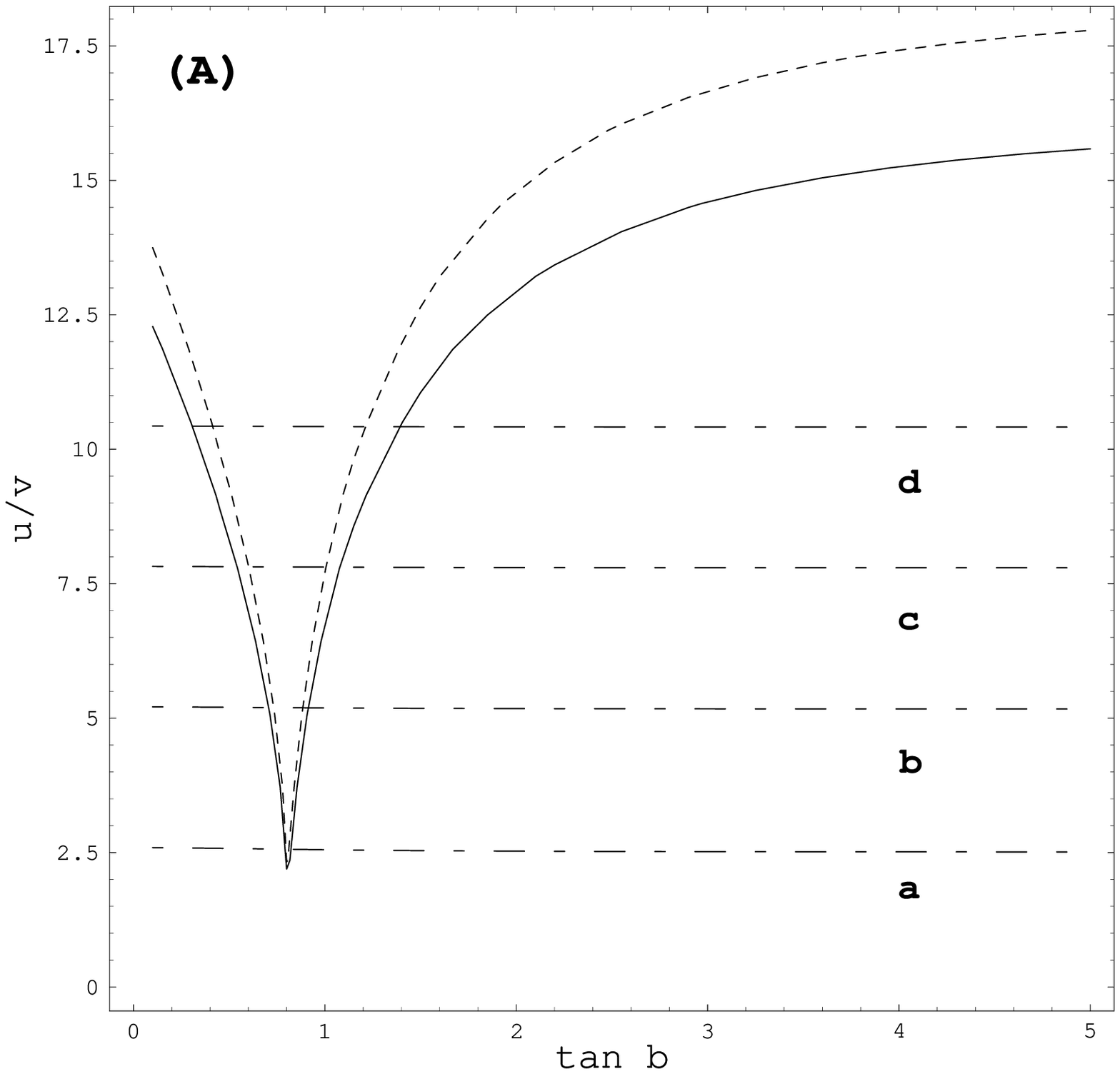}
\hspace*{1cm}
\epsfysize=7cm
\leavevmode
\epsfbox{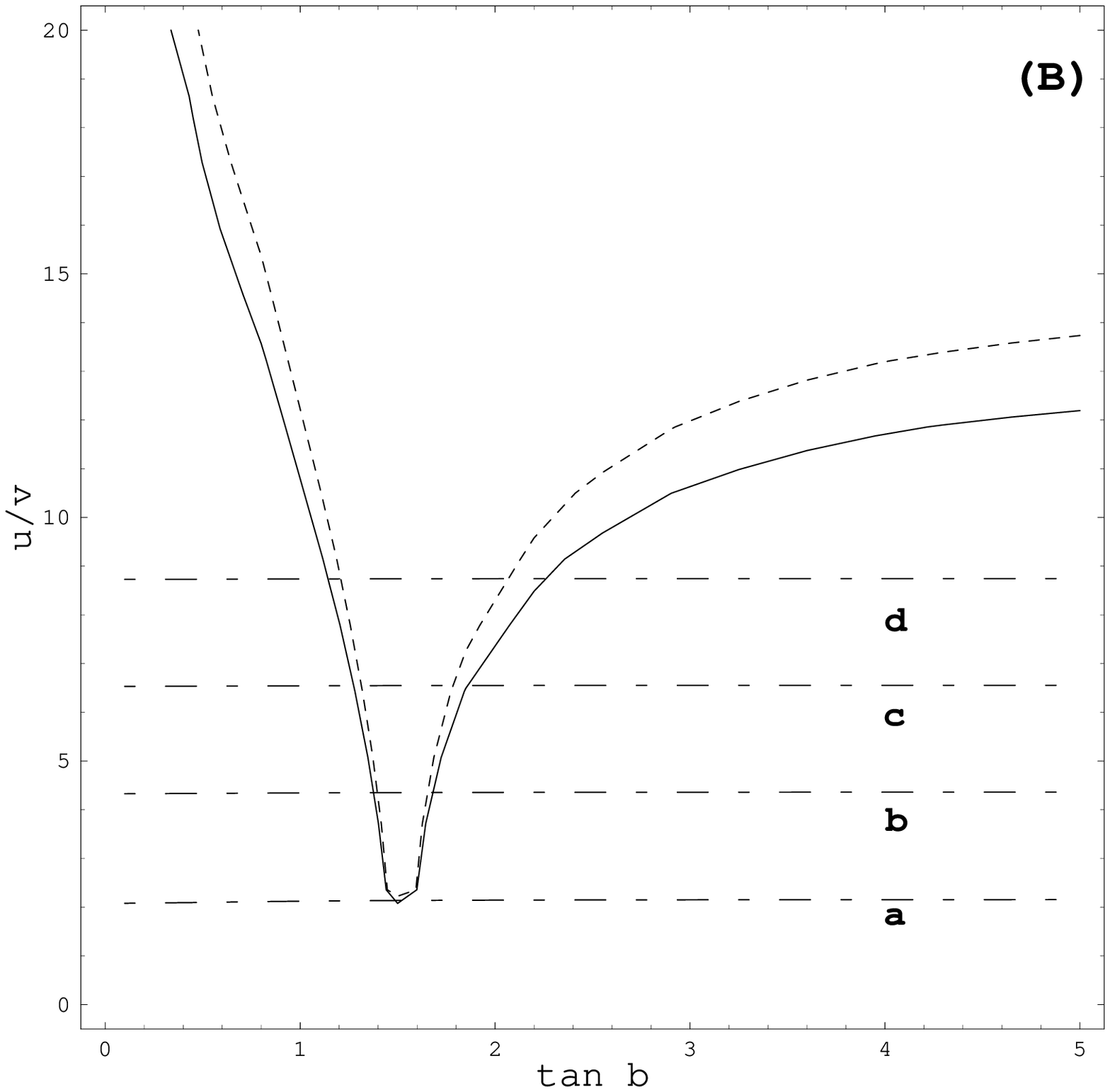}
\end{center}
\vspace*{-3mm}
{\footnotesize Fig.2\  $\chi^2$-contours in the $(\tan\beta, ~u/v)$ plane
for (A) $\eta$ model with $\sin\chi=0.3$ and (B) $\xi_-$ model with
$\sin\chi=0.2$. In both cases $g_X/g_Y=1$ is assumed. 
Contours are shown by a solid line for $\chi^2=18.5$ and a dotted line for 
$\chi^2=14.1$. 
Dashed-Dotted lines represent the $m_{Z_2}$ contours a: 400 GeV, b:
800 GeV, c: 1200 GeV and d: 1600 GeV, respectively.}
\end{figure}

A gross shape of the $\chi^2$-contours shown in Fig.2 is 
general and they donot show 
so strong $g_X$ and $\sin\chi$ dependence.
However, the value of $\tan\beta$ allowed at the small $u/v$ region 
somehow changes depending on the 
$\sin\chi$ value as expected from Eq. (41). 
If we note a $\chi^2 =18.5$ contour, we find that the region of 
$u/v \le 10$ can be realized at very restricted $\tan\beta$ values as follows,
\begin{eqnarray}
\eta~~{\rm model}&& \left\{ 
\begin{array}{ll}
0<\tan\beta~{^<_\sim}~1.1 & (~\sin\chi=0~), \\
0.4~{^<_\sim}~\tan\beta~{^<_\sim}~1.2 & (~\sin\chi=0.3~), 
\end{array} \right.
\nonumber \\
\xi_-~{\rm model}&& \left\{
\begin{array}{ll}
0.9<\tan\beta~{^<_\sim}~1.9 & (~\sin\chi=0~), \\
1.2<\tan\beta~{^<_\sim}~2.6 & (~\sin\chi=0.2~). 
\end{array} \right. 
\end{eqnarray}
In both models the somehow larger $\sin\chi (\ge 0)$ value 
is necessary to realize $\tan\beta$
in a favorable region like $\tan\beta~{^>_\sim}~1$ 
as far as we require that rather small $u/v$ value is allowable.
Generally, in the $\xi_-$ model the naturally small 
$\sin\chi (\ge 0)$ values can make $u/v$ rather small 
for the $\tan\beta~{^>_\sim}~1$ region.
Thus the $\xi_-$ model may be more promising than the $\eta$ 
model from this viewpoint.  

\subsection{Minimization of scalar potential} 
Next we study the vacuum structure through minimizing
the scalar potential \cite{extra1,cdeem,wl}.
We again take account of the influence on the scalar potential
caused by the kinetic term mixing among abelian gauge fields.
It should be reminded that the abelian auxiliary fields $D_{Y,X}$ 
are changed as shown in Eq. (21).
For the detailed investigation of the scalar potential in the present
models the superpotential and the soft supersymmetry breaking terms
should be fixed definitely.
In this paper we assume the following superpotential,
\begin{equation}
W=h_UQH_2\bar U+h_DQH_1\bar D+h_ELH_1\bar E +h_NLH_2\bar N 
+ \lambda SH_1H_2 + kS\bar gg. 
\end{equation}
Here we explicitly write the minimal part which is necessary for this model
to be realistic. 
Generation indices are abbreviated.
Other terms including the exotic fields are
omitted.\footnote{We also drop the usual R-parity violating terms to
guarantee the proton stability. This may be justified due to some
discrete symmetry. The last term is necessary for the radiative
symmetry breaking of U(1)$_X$ as discussed in \cite{extra1,wl}.
In the $\eta$ model $h_N=0$ should be assumed.}
As soft supersymmetry breaking terms we consider the following
ones,\footnote{We donot consider the abelian gaugino mass mixing term
$M_{YX}\lambda_Y\lambda_X$, for simplicity.} 
\begin{eqnarray}
{\cal L}_{\rm soft}&=&-\sum_km_k^2\vert\phi_k\vert^2-
m_1^2\vert H_1\vert^2 -m_2^2\vert H_2\vert^2
-m_S^2\vert S\vert^2 \nonumber \\ 
&+&\Bigg[ A_Uh_UQH_2\bar U+A_Dh_DQH_1\bar D +A_Eh_ELH_1\bar E 
+ A_Nh_NLH_2 N \nonumber \\ 
&&\hspace*{5mm}+A_\lambda \lambda SH_1H_2 +A_k k S\bar gg \nonumber \\
&&\hspace*{5mm}+ {1 \over 2}\left(M_W\sum_i\lambda_{W_i}\lambda_{W_i} 
+ M_Y\lambda_Y\lambda_Y +M_X\lambda_X\lambda_X\right) + h.c. \Bigg],
\end{eqnarray}
where $\phi_k$ stands for the superpartners of ordinary quarks and leptons.
For simplicity, $A$-parameters and Yukawa couplings are assumed to be real.

The changes induced in the scalar potential 
by the extra U(1)$_X$ is expected to be reflected in the vacuum structure.
In order to investigate this aspect we write down
the vacuum energy by using the VEVs of $H_1,~ H_2$ and $S$.
Under the assumption (32) for these VEVs,
we can write it as follows,
\begin{eqnarray}
V&=&{1 \over 8}\left(g_W^2+g_Y^2\right)\left(v_1^2-v_2^2\right)^2\nonumber\\
&&+{1 \over 8}\left\{g_Y\tan\chi(v_1^2-v_2^2)+{g_X \over \cos\chi}
\left( Q_1(v_1^2-u^2)+Q_2(v_2^2-u^2)\right)\right\}^2 \nonumber \\
&&+\lambda^2v_1^2v_2^2+\lambda^2u^2v_1^2+\lambda^2u^2v_2^2 
+m_1^2v_1^2 +m_2^2v_2^2+m_S^2u^2 -2A\lambda uv_1v_2.
\end{eqnarray}
Here we use the relation $Q_1+Q_2+Q_S=0$ for the extra U(1)$_X$ charge, 
which comes from the $\lambda SH_1H_2$ coupling in the superpotential $W$.
In the second line we can see the effect of the kinetic term mixing.
A linear term of $u$ appears in this potential only in a form $Au$ so
that a sign of $A$ is relevant to the one of $u$.

Now we proceed an analysis of the feature of the minimum of this 
potential.
The analytic study is difficult and then 
the potential minimum must be numerically studied by solving the 
minimum conditions,
\begin{equation}
{\partial V\over \partial v_1}={\partial V\over \partial v_2}
={\partial V\over \partial u}=0.
\end{equation}
These conditions can be translated into the equations for Higgs masses,
\begin{eqnarray}
\left({m_1\over v}\right)^2&=&
{g_W^2+g_Y^2\over 4}\left( \sin^2\beta-\cos^2\beta\right)
-{\zeta_1\over 4}\left(\zeta_1\cos^2\beta
+\zeta_2\sin^2\beta+\zeta_3\left({u\over v}\right)^2\right) \nonumber \\ 
&&-\lambda^2\left(\sin^2\beta+\left({u\over v}\right)^2\right)
+\lambda{A\over v}{u\over v}\tan\beta,\\
\left({m_2\over v}\right)^2&=&
-{g_W^2+g_Y^2\over 4}\left( \sin^2\beta-\cos^2\beta\right)
-{\zeta_2\over 4}\left(\zeta_1\cos^2\beta+\zeta_2\sin^2\beta
+\zeta_3\left({u\over v}\right)^2\right)
\nonumber \\
&&-\lambda^2\left(\cos^2\beta+\left({u\over v}\right)^2\right)
+\lambda{A\over v}{u\over v} \cot\beta, \\
\left({m_S\over v}\right)^2&=&-{\zeta_3\over 4}
\left(\zeta_1\cos^2\beta+\zeta_2\sin^2\beta
+\zeta_3\left({u\over v}\right)^2\right)
-\lambda^2 +\lambda {A\over v}{v \over u}
\cos\beta\sin\beta,
\end{eqnarray}
where $v_1^2+v_2^2=v^2$ should be satisfied.
We define $\zeta_1, ~\zeta_2$ and $\zeta_3$ as
\begin{equation}
\zeta_1=g_Y\tan\chi+{g_XQ_1\over\cos\chi}, \quad 
\zeta_2=-g_Y\tan\chi+{g_XQ_2\over\cos\chi},\quad
\zeta_3={g_XQ_S\over\cos\chi}.
\end{equation}
A solution of Eqs. (58)-(60) is realized as a triple crossing
point of the contours of $m_1^2,~ m_2^2$ and $m_S^2$ in the
$(\tan\beta, ~u/v)$ plane for a suitable set of parameters
$g_X$, $\sin\chi$, $A$ and $\lambda$. 
It is not an easy task to solve the coupled RGEs for all physical
parameters and find solutions of Eqs. (58)-(60) varying the initial conditions 
for all parameters at the unification scale.
In the present analysis we donot practice this procedure 
but adopt more convenient method.

To know the existence of such solutions we should search the parameters
region for which the contour bands of $m_1^2,~ m_2^2$ and 
$m_S^2$ simultaneously cross each other within a suitable width like 
$\vert m^2\vert ~{^<_\sim}~(1~{\rm TeV})^2$.
The absolute values of these squared masses are naturally considered
to be near the weak scale from a viewpoint of the radiative symmetry 
breaking induced by the quantum corrections to the soft supersymmetry 
breaking parameters.
These parameters are usually considered to be from a few handred GeV
to 1 TeV at the unification scale and run towards the low energy region mainly 
under the control of the contribution from large Yukawa coupling constants.
If we take this viewpoint, we can roughly know the consistent 
parameter region with the radiative symmetry breaking scenario.
We follow this simplified method.
The complete RGEs study done in \cite{extra1,cdeem,wl} seems to support
the result of this method.

\begin{figure}[htb]
\begin{center}
\epsfysize=7cm
\leavevmode
\epsfbox{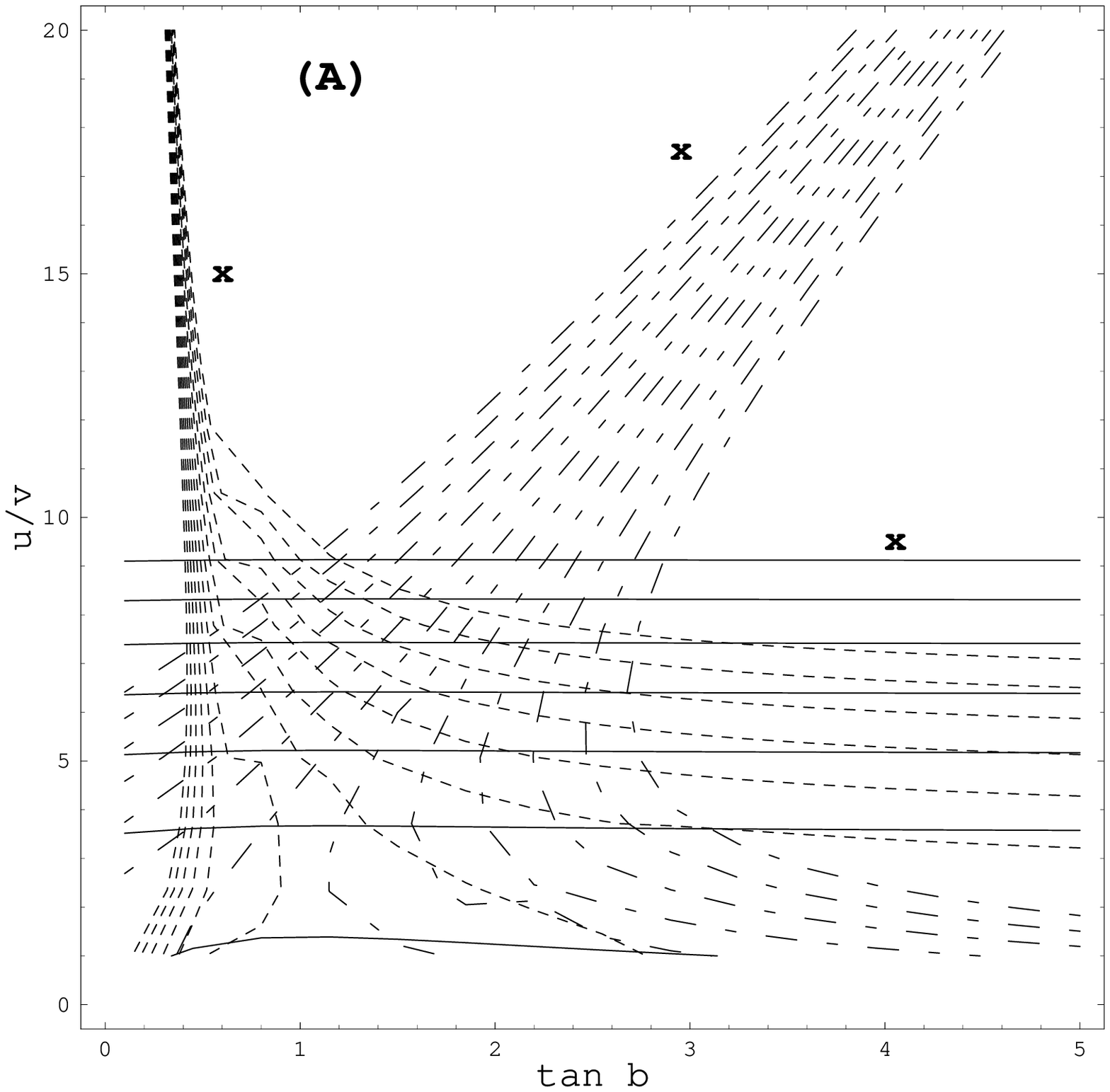}
\hspace*{1cm}
\epsfysize=7cm
\leavevmode
\epsfbox{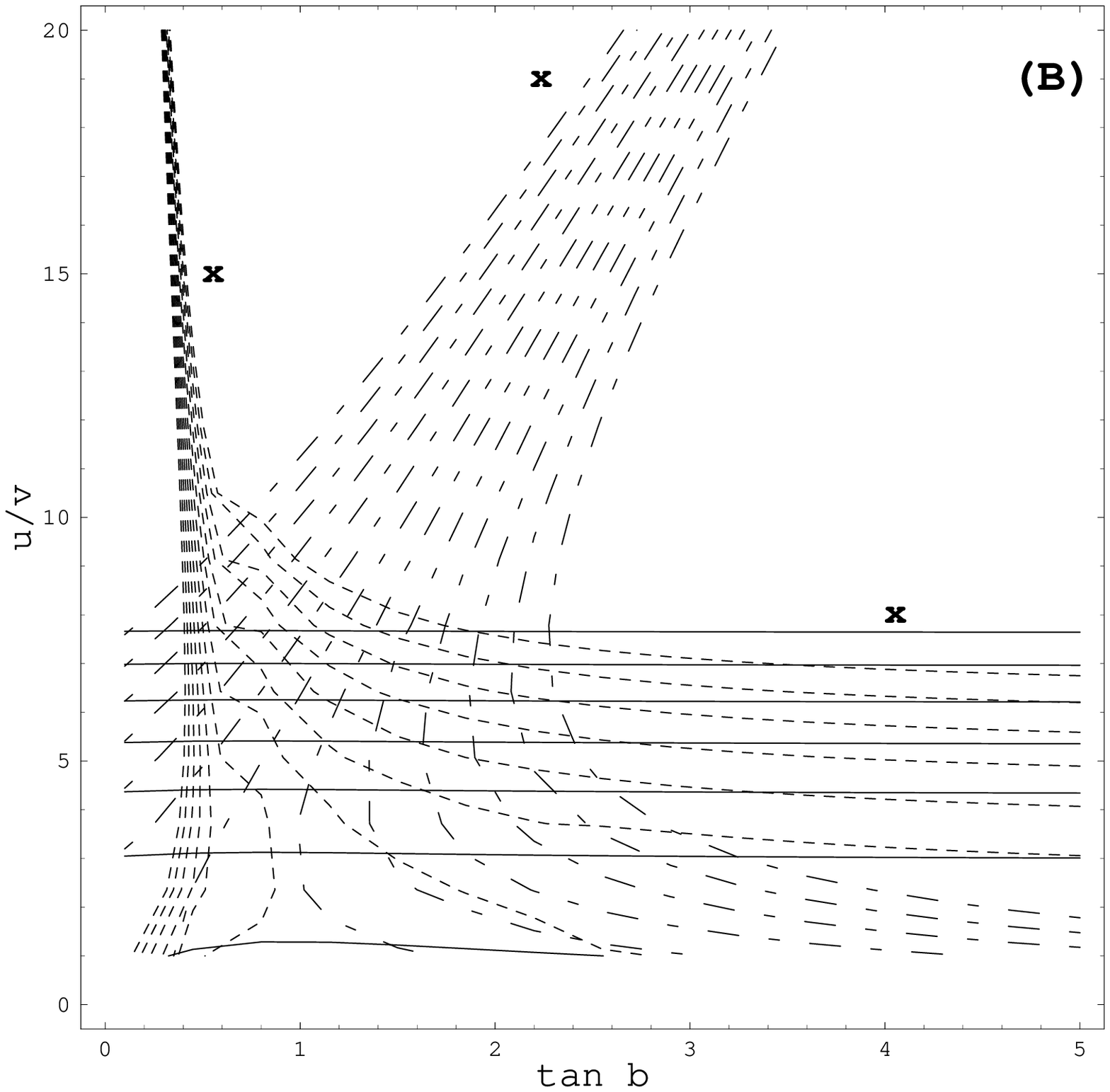}
\end{center}
\vspace*{-3mm}
{\footnotesize Fig.3\ Contours of $m_1^2$, $m_2^2$ and $m_S^2$  
in the $(\tan\beta, ~u/v)$ plane
for (A) $\eta$ model with $\sin\chi=0.3$ and (B) $\xi_-$ model
with $\sin\chi=0.2$. In both cases $g_X/g_Y=1$, $\lambda=0.5$ and
$A/v=2$ are assumed.
Contours of $m_1^2$, $m_2^2$ and $m_S^2$ are drawn by
dash-dotted, dotted and solid lines, respectively. 
The value associated with contours are $(m_{1,2,S}/v)^2=
-30,-25,-20,\cdots,20,30$. In particular,
contours for $(m_{1,2,S}/v)^2=-30$ are marked by x at their neighborhood. }
\end{figure}

We numerically examine the possible vacuum structure by drawing 
$m_1^2$, $m_2^2$ and $m_S^2$ contours 
in the $(\tan\beta, ~u/v)$ plane for the various values of parameters $g_X$,
$\sin\chi$, $g_X$, $\lambda$ and $A$.
We vary these in the region (52) and
\begin{equation}
0.1\le \lambda\le 0.9, \qquad \vert A/v \vert \le 10.
\end{equation} 
Since the parameter $A$ is varied in the above region including a
negative $A$, 
we only need to search the positive $u/v$ region.
As such an example, in Fig. 3 we draw those contours 
for a typical set of these parameters.

Here we summarize the general features found from this analysis.
The $\sin\chi$ and $g_X$ dependence of these contours is
non-negligible but weak.
As expected from the largeness of $u/v$, 
$m_S^2$ is almost dominated by the first term in Eq. (60) and then it is 
not affected so large by parameters $\lambda$ and $A$ in our
interesting $(\tan\beta, ~u/v)$ region. 
The contours of $m_S^2$ are almost equivalent to the constant $u/v$ 
lines satisfying $u/v ~{^<_\sim}~10$.
On the other hand, parameters $\lambda$ and $A$ play crucial role to
determine the contours $m_1^2$ and $m_2^2$.
This feature is mainly related to the behavior of the last terms in
Eqs. (58) and (59).
The behavior of contours $m_1^2$ and $m_2^2$ in our considering
$(\tan\beta,~u/v)$ regime may be explained as follows.
For the small $\lambda$ and $\vert A/v\vert$ values, the $m_1^2$ and 
$m_2^2$ contours are also almost constant $u/v$ lines since the first
two terms in Eqs. (58) and (59) dominate their values.
As its result there appears an allowed wide $\tan\beta$ region where 
three contours satisfying $\vert (m/v)^2\vert < 30 $ can cross at a point.
If we take $\vert A/v\vert$ larger, the overlapping region of three 
contours shrinks around $\tan\beta\sim1$ and also its upper bound 
on $u/v$ becomes smaller.
This is because the contours of $m_1^2$ and $m_2^2$ have a shape which
is determined by the last terms of Eqs. (58) and (59). 
If we make $\lambda$ larger, their shape does not change largely.
Although a wide $\tan\beta$ region gives a solution and no favorite 
$\tan\beta$ region appears, the upper bound on $u/v$ becomes smaller.
When both of $\lambda$ and $\vert A/v\vert$ are taken larger, the contours of
$m_1^2$ and $m_2^2$ cross at the region where $u/v$ has too 
large or small values like $u/v~{^>_\sim}~10$ or $u/v~{^<_\sim}~1$.
In that case we cannot find a solution in a suitable $(m_S/v)^2$
region. Some of these fetures can be found in Fig. 3.
Similar features are reported in the RGEs study \cite{wl}.

These results also may give us a hint for the RGEs study. 
As found in this argument, in such a study one of the
important problems is clearly to find what kind of input parameters can
realize the negative smaller $m_S^2$ at the weak scale to obtain such 
a solution as $u/v~{^>_\sim}~2$.
In that case we may need non-universal soft supersymmetry breaking
scalar masses at least in the Higgs sector \cite{extra1,wl}.

Now by combining this result with the previously discussed
$\chi^2$-fits using the
precise measurements, we can restrict the allowed region in the 
$(\tan\beta, ~u/v)$ space of the present models.
In that region the model can simultaneously satisfy both of
the condition for the radiative symmetry breaking and 
the requirement to fulfill the constraint from the data of 
electroweak precise measurements.
As an example, if we use Figs. 2 and 3 for this purpose, 
we can find such solutions at the following values,
\begin{eqnarray}
&& \tan\beta\sim 0.8, \qquad 2.5~{^<_\sim}~u/v~{^<_\sim}~ 8 \qquad 
(~\eta~{\rm model~with}~\sin\chi=0.3~), 
\nonumber \\ 
&& \tan\beta\sim 1.5, \qquad 2~{^<_\sim}~u/v~{^<_\sim}~8 \qquad (~\xi_-~{\rm
model~with}~\sin\chi=0.2~ ), 
\end{eqnarray} 
where the parameters are taken as $g_X/g_Y=1,~\lambda=0.5$ 
and $A/v=2$.
If we deviate $\tan\beta$ from the above one within the
region (53), the lower bound of $u/v$ suddenly becomes larger.
These are general features.

We comment in some detail on the parameters dependence of these solutions.
The drastic effect of $\sin\chi\not= 0$ on this vacuum 
structure can not be found but we should note that 
it has an important role to determine $\tan\beta$ which 
makes small $u$ values allowable.
Generally the consistent solutions tends to be found for not so large 
values of $\lambda$ and $\vert A/v\vert$. 
This tendency can be understood from the previously mentioned
features of the parameter dependence of Fig. 2 and Fig. 3.
If we take account of the well known result $\tan\beta~{^>_\sim}~1$ in the 
study of the radiative symmetry breaking due to large Yukawa couplings, 
the above results shows the $\eta$ model seems not to be realized as a
consistent model in our framework assuming $\vert
(m_{1,2,S}/v)^2\vert \le 30$.
The $\xi_-$ model has a good nature also in this aspect.
 
\subsection{Masses of an extra $Z$ and a neutral Higgs scalar}
It is very interesting that in this study 
the vacuum solutions can be found only at the stringently
restricted region in the $(\tan\beta, ~u/v)$ plane.
This makes us possible to find the bounds on $m_{Z_2}$
and the lightest neutral Higgs scalar mass $m_{h^0}$.
Here we estimate these mass bounds in the basis of the study in the
previous subsections. 

The neutral Higgs scalars mass matrix can be written in the basis of
$(H_1^0,~H_2^0, ~S)$ by using the minimization 
conditions (58)-(60) as,

\footnotesize
$$
{m_Z^2 \over \tilde g^2}  
\left( \begin{array}{ccc}
\cos^2\beta(\tilde g^2 + \zeta_1^2)+ 2 \lambda \tilde A \tilde u\tan\beta&
{1\over 2}\sin 2\beta(-\tilde g^2 + X_{12})
 -2 \lambda \tilde A \tilde u & 
\cos\beta\left(\tilde u X_{13}
 - 2\lambda \tilde A\tan\beta\right)\\
{1\over 2}\sin 2\beta(-\tilde g^2 + X_{12})
 -2 \lambda \tilde A \tilde u&
 \sin^2\beta(\tilde g^2 +\zeta_2^2)+ 2\lambda \tilde A \tilde u\cot\beta&
\sin\beta\left(\tilde u X_{23} - 2\lambda\tilde A\cot\beta\right) \\
\cos\beta\left(\tilde u X_{13}- 2 \lambda\tilde A\tan\beta\right)&
\sin\beta\left(\tilde u X_{23}
 -2\lambda\tilde A \cot\beta\right)&
\zeta_3^2\tilde u^2+{\tilde A \over \tilde u}\lambda \sin 2\beta\\
\end{array}\right), 
$$
\normalsize

\noindent
where $\tilde g^2=g_W^2+g_Y^2$ and $X_{ij}=\zeta_i\zeta_j+4\lambda^2$.
In this expression we also use abbreviations $\tilde A=A/v$ and $\tilde u=u/v$.
In order to estimate $m_{h^0}$
we numerically diagonalize this matrix and plot the contours of the 
smallest mass eigenvalue in the $(\tan\beta, ~u/v)$ plane.
Since in the present models the value of $u/v$ is severely 
restricted by the bounds on 
$m_{Z_2}$ and $\xi$, it cannot be so small that the lightest neutral
Higgs scalar 
cannot be generally dominated by a scalar partner of $S$ which has no
electroweak interactions. 
This situation is completely different from the case of NMSSM
where the lightest neutral Higgs scalar can be dominated by the scalar
component of $S$ in the case of the small $u/v$ \cite{nmssm}.
Thus the MSSM bound on the lightest neutral Higgs scalar mass might be almost 
applicable and we could impose $m_{h^0}~{^>_\sim}~62.5$~GeV \cite{particle}.
This can give an additional constraint on the parameters in the present
models.

\begin{figure}[htb]
\begin{center}
\epsfysize=7cm
\leavevmode
\epsfbox{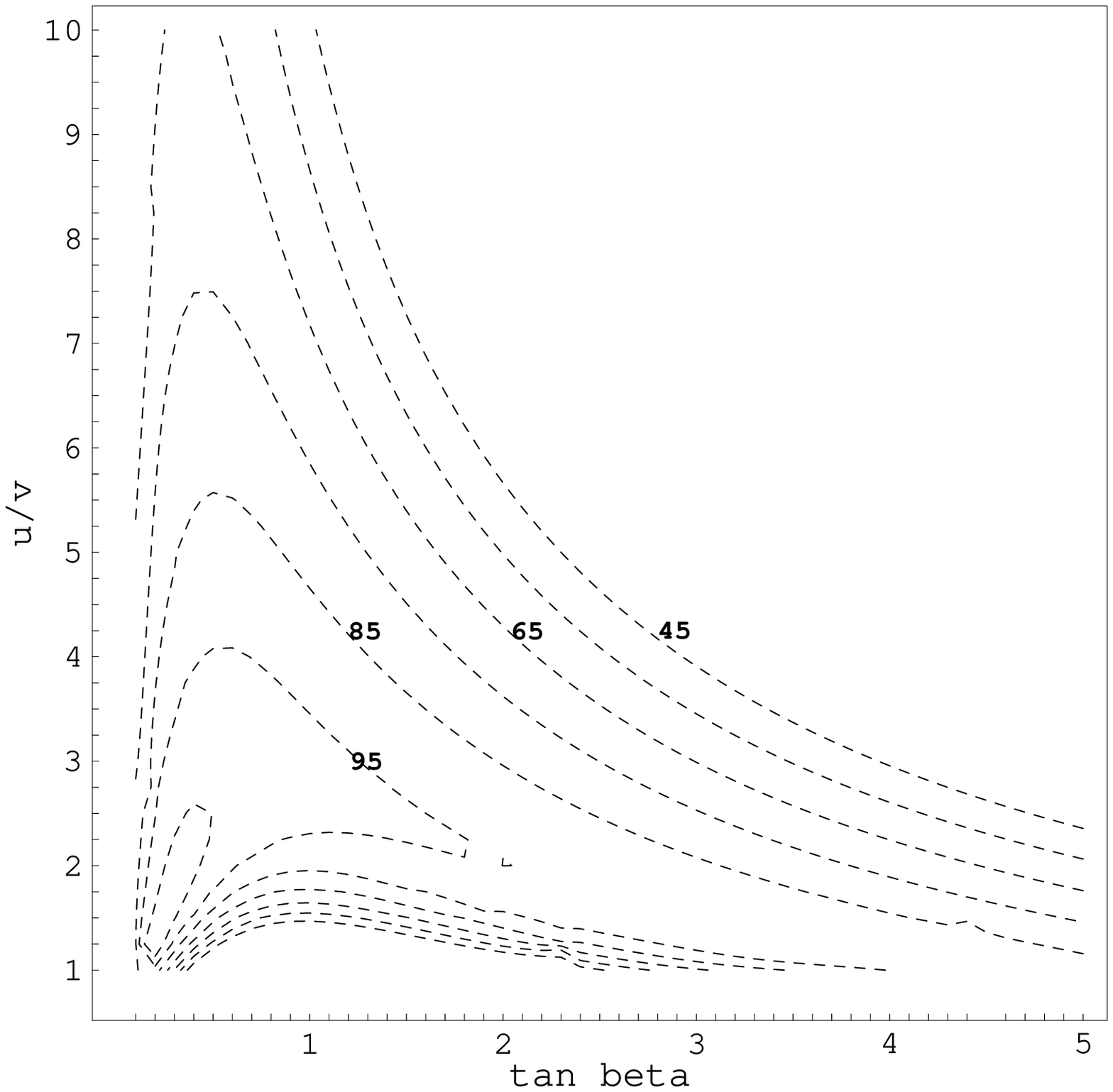}
\end{center}
\vspace*{-3mm}
{\footnotesize Fig.4\ Contours of the lightest neutral Higgs scalar mass
$m_{h^0}$ in the $(\tan\beta, ~u/v)$ plane
for $\xi_-$ model with $\sin\chi=0.2,~g_X/g_Y=1,~\lambda=0.5$ and
$A/v=2$.} 
\end{figure}

To make our discussion definite we consider the $\xi_-$ model.
We know from the discussions up to now 
that $\tan\beta~{^>_\sim}~1$ and $2~{^<_\sim}~u/v~{^<_\sim}~8$ are required.
This makes us possible to examine the allowed parameters region 
by checking whether this condition on $m_{h^0}$ is satisfied or not
in the constrained $(\tan\beta, ~u/v)$ plane.
Although $m_{h^0}$ is found to have only a negligible dependence 
on $\sin\chi$ and $g_X$, it is crucially dependent on $\lambda$ and $A$.
In fact, for $A/v~{^<_\sim}~0.1$, $m_{h^0}~{^>_\sim}~62.5$~GeV cannot be
satisfied in the favorable part of the $(\tan\beta, ~u/v)$ plane 
for any $\lambda$ in the region (62).
For small $A/v$, the larger $\lambda$ has a tendency to make 
$m_{h^0}^2<0$ there and then the vacuum unstable.
We can estimate the allowed perturbative region of $\lambda$ 
for the typical values of $A/v$, for which $m_{h^0}$ can satisfy the
experimental lower bound in a certain point of the above mentioned
constrained $(\tan\beta, ~u/v)$ plane.
Such an example is
\begin{equation}
\begin{array}{ll}
0.2~{^<_\sim}~\lambda~{^<_\sim}~0.5 &\qquad {\rm for}~~A/v=1, \\
0.4~{^<_\sim}~\lambda~{^<_\sim}~0.9 &\qquad {\rm for}~~A/v=5. 
\end{array}
\end{equation}
For the larger $A/v$, the larger $\lambda$ is favored and the larger
$A/v$ tends to make $m_{h^0}$ larger. 

To find the absolute value of $m_{h^0}$ consistent with other conditions, we 
take the same parameters as those used to draw Fig. 2 and Fig. 3.
In this case contours of the lightest neutral Higgs scalar mass 
are shown in Fig. 4. 
One of interesting features of this figure may be a behavior of the
contours for $\tan\beta$.
As mentioned before, the lightest Higgs scalar in this model 
is mostly composed in
the same way as the one of the MSSM whose mass increases with increasing
$\tan\beta$. On the other hand, Fig. 4 shows a completely different
behavior from that.\footnote{The author would like to thank the referee
for pointing this out.}
In the present Higgs scalar mass matrix there is a D-term contribution
of the extra U(1)$_X$ even in a 2$\times 2$ submatrix for $(H_1^0,
H_2^0)$.
This contribution seems to make us possible to understand the feature
in Fig. 4.
In order to see this briefly, we may extract a 2$\times 2$ submatrix
for $(H_1^0, H_2^0)$ and carry out orthogonal transformation with an
angle $\beta$.
Then we can find the lightest Higgs mass upper bound as \cite{drees,extra1}
$$
m_{h^0}^2 \le m_Z^2\left[ \cos^4\beta\left(1+{\zeta_1^2\over \tilde 
g^2}\right)+\sin^4\beta\left(1+{\zeta_2^2\over \tilde 
g^2}\right)+2\sin^2\beta\cos^2\beta\left(-1+{X_{12}\over 
\tilde g^2}\right)\right],
$$
where $\zeta_1,~\zeta_2$ and $X_{12}$ represent the extra U(1)$_X$
effect. 
Although this upper bound increases with increasing $\tan\beta$ in the 
case of $\zeta_1=\zeta_2=0$,
we can find that this can decrease with increasing $\tan\beta$
in certain extra U(1)$_X$ models like the present case.
The extra U(1)$_X$ model may give an example of the lightest 
Higgs scalar with the similar composition to the MSSM one
but with the different mass behavior for $\tan\beta$ from that.

Finally, if we combine this with Figs. 2 and 3, 
the bounds of $m_{Z_2}$ and $m_{h^0}$ can be estimated as
\begin{equation}
400~{\rm GeV}~{^<_\sim}~m_{Z_2}~{^<_\sim}~1500~{\rm GeV} , \qquad
62.5~{\rm GeV}~{^<_\sim}~m_{h^0}~{^<_\sim}~95~{\rm GV}.
\end{equation}
In the case of $m_{Z_2}$ the lower and upper bounds are mainly
constrained  by the values of $\sin\chi$ and $m_{1,2,S}^2$, respectively.
On the other hand, we should note again that $\lambda$ and $A$ are the 
most important to determine the bound of $m_{h^0}$. 
Although we assume that the soft supersymmetry breaking scalar masses $m_1^2$,
$m_2^2$ and $m_S^2$ should be in the reasonable range
$\vert m^2\vert <(1~{\rm TeV})^2$ from the viewpoint of naturalness 
consideration, it seems to be difficult to change 
this condition largely so that we expect our estimation method and
then the obtained results are not so bad as far as $\lambda$ and $A$
are fixed in a suitable range.

\section{Summary}
The weak scale extra U(1)$_X$ models are very promising candidates of 
a solution for the $\mu$-problem.
They are characterized by the existence of an additional neutral gauge boson
around a several handred GeV or a TeV.
Thus they are also interesting from a viewpoint of experiments at the
high energy front.
As such a concrete example, at least we have two simple models, $\eta$ model 
and $\xi_-$ model which can be induced from the perturbative 
superstring inspired $E_6$ model. 
For these models it is an important issue to check their consistency 
as realistic ones.

In this paper we payed our attention on their vacuum structure
parametrized by the VEVs of three Higgs scalars to study it.
We investigated them based on the constraints from the precise measurements 
of electroweak observables and the radiative symmetry breaking at the weak 
scale.
We searched the allowed region in the $(\tan\beta, ~u/v)$ plane.
When we proceeded this study, we took account of the effects of 
the gauge kinetic term mixing of abelian factor groups, which might be 
a special feature of the extra U(1) models.
In general they can potentially bring some effects not only on the 
usual gauge interaction sector but also on the D-term scalar
potential. The latter effect has not been payed any attention before.

Based on the numerical study we had the following results.
First, the consistent region in the $(\tan\beta, ~u/v)$ plane is strongly
restricted but we could find that there was the interesting 
allowed region for suitable values of parameters $\lambda$ and $A$. 
This means that the weak scale extra U(1)$_X$ models could be realistic
as the $\mu$-problem solvable electroweak model at least from the 
viewpoint of consistency of the precise measurements and the radiative
symmetry breaking. 
Using our result for the allowed region in the $(\tan\beta, ~u/v)$
plane, we could estimate the mass bounds on an extra neutral gauge boson and
the lightest neutral Higgs scalar. It is noticable that we can 
bring the upper bound on the lightest
neutral Higgs scalar mass.
Since we didnot solve the RGEs of physical parameters like Yukawa
coupling constants, soft supersymmetry breaking parameters and so on 
explicitly in this analysis, these bounds could be somehow rough ones.
However, it is an interesting feature of this type of analysis to 
put the upper bound for them.
To obtain more fine bounds, we need to analyze it by solving RGEs explicitly.

Second, the effect on the scalar potential itself coming from the 
kinetic term mixing was found not to be so large.
However, its effect could not be negligible
as it could rather largely affect the gauge interaction sector.
Since the allowed region in the $(\tan\beta, ~u/v)$ plane is
influenced by both sectors, we could see non-negligible effects on the
values of $\tan\beta$ and $u$ due to the kinetic term mixing. 
We need further quantitative study for the origin of this kinetic term 
mixing in the present type of models as done in \cite{mix2,riz}.

The $\mu$-problem solvable models, in particular, which have an intermediate
scale like our $\xi_-$ model, show very interesting phenomenological
features such as an explanation of small neutrino masses. 
Although we only discuss the $\xi_-$ model as an example of 
this kind of models, 
there can be other models and it seems to be worthy to construct 
such models by various methods, especially, as the low energy effective 
models of superstring. 
If we consider these models seriously, we also need to study the problems
related to the exotic fields included in the models, 
like proton stubility and flavor changing rare processes \cite{sue0}.
Such a study may enlarge our scope for the unification scenario
different from the usual grand unification schemes.

\vspace*{7mm}
The author would like to thank M.~Konmura, M.~Itoh and Tadao Suzuki for the
collaboration at the first stage.
He also thanks the referee for useful and important comments on the manuscript 
to improve its content. 
This work has been supported in part 
by a Grant-in-Aid for Scientific Research from the Ministry 
of Education, Science and Culture(\#08640362).

\newpage
{\Large\bf Appendix}\vspace{5mm}

In this appendix we derive the formulae for 
the deviation of the electroweak parameters
($\rho$, $\bar s_W^2$) from the SM prediction due to the existence 
of an extra $U(1)_X$ gauge field based on Refs. \cite{electro,kinet}. 
They are defined in the SM through 
the neutral current effective interaction Lagrangian as follows,
\begin{equation}
{\cal L}_{\rm int}={e \over 2 s_Wc_W}
\rho^{1/2}_{\rm SM}\left(J_3^\mu-2\bar s_W^2J_{\rm em}^\mu\right)Z_\mu,
\end{equation}
where $\rho_{\rm SM} \simeq 1+\delta\rho_t$.
The corresponding effective interaction Lagrangian of $Z_{1\mu}$ in the 
extra $U(1)_X$ model is defined as
\begin{equation}
{\cal L}_{\rm int}=
{e \over 2s_Zc_Z}\rho^{1/2}(J_3^\mu -2\bar
s_Z^2J_{\rm em}^\mu)Z_{1\mu}.
\end{equation}
We assume that the radiative correction dominantly comes from
the SM sector and others can be neglected.
Thus the one-loop corrected on-shell Weinberg angle $\bar s_W^2$ in the SM
and the corresponding Weinberg angle $\bar s_Z^2$ in the present models
are defined in terms of the observed values $\alpha$ and $G_F$ as
\begin{equation}
\bar s^2_W\bar c^2_W={\pi\alpha(m_Z^2)\over\sqrt 2G_Fm_Z^2
\rho_{\rm SM}}, \qquad
\bar s^2_{Z}\bar c^2_{Z}={\pi\alpha(m_{Z}^2)\over\sqrt
2G_Fm_{Z_1}^2\rho_{\rm SM}},
\end{equation}
where the relation between $m_{Z_1}^2$ and $m_Z^2$ is given by
Eq. (42).
From Eq. (68) we can easily obtain
\begin{equation}
\bar s_{Z}^2\simeq \bar s^2_W+ {\bar s^2_W \bar c^2_W\over \bar 
c^2_W- \bar s^2_W} \left({m_Z^2\over m_{Z_1}^2}-1 \right).
\end{equation}

In the present model vector interaction parts in the effective Lagrangian 
for $Z_{1\mu}$ are summarized by using Eqs. (44) and (45) as follows,
\begin{eqnarray}
&&{\cal L}_{\rm int}={e \over 2s_Wc_W}\rho_{\rm SM}^{1/2}
\left( 1 +s_W\xi\tan\chi\right)
\left(J_3^\mu-2\bar s^2_ZJ_{\rm em}^\mu\right)Z_{1\mu} \nonumber \\
&&\qquad -ec_W\xi\tan\chi J_{\rm em}^\mu Z_{1\mu} 
+{g_X\xi \over 2\cos\chi}J^{\prime\mu} Z_{1\mu}.
\end{eqnarray}
The last term is taken into account as a $\bar v_f^\prime$ term in
Eq. (45) and then it is irrelevant to the present calculation.
The first two terms in the right-hand side can be rearranged into 
the following form,
\begin{equation}
{\cal L}_{\rm int}\simeq {e \over 2s_Wc_W}\rho_{\rm SM}^{1/2}
\left( 1 +s_W\xi\tan\chi\right)
\left(J_3^\mu-2\bar s^{\prime 2}_Z J_{\rm em}^\mu\right)Z_{1\mu}.
\end{equation}
Since the $Z_\mu$-$X_\mu$ mixing introduces a new interaction
$-ec_W\xi\tan\chi J_{\rm em}^\mu Z_{1\mu}$ for $Z_{1\mu}$
at tree level in comparison with the SM case,
 $\bar s_Z^{\prime 2}$ deviates from $\bar s_Z^2$ as
\begin{equation}
\bar s_Z^{\prime 2}\simeq \bar s_Z^2+ s_Wc^2_W\xi\tan\chi.
\end{equation}
We can also rewrite the right-hand side of Eq. (71) as
\begin{equation}
{e \over 2s_{Z}c_{Z}}\rho_{\rm SM}^{1/2}\left( 1
+s_W\xi\tan\chi\right)
\left[ 1+{1\over 2}\left({m_Z^2 \over m_{Z_1}^2}-1 \right)\right]
\left(J_3^\mu-2\bar s^{\prime 2}_Z J_{\rm em}^\mu\right)Z_{1\mu}.
\end{equation}
From this expression we can extract the expression of $\rho$ parameter
in this model,
\begin{equation}
\rho=\rho_{\rm SM}(1+s_W\xi\tan\chi)^2
\left[ 1+{1 \over 2}\left({m_Z^2 \over m_{Z_1}^2}-1 \right)\right]^2.
\end{equation} 
Thus up to the first order of the small quantities
the expression of $\delta\rho_M$ can be read off as,
\begin{equation}
\delta\rho_M\simeq 2s_W\xi\tan\chi+\left({m_Z^2\over m_{Z_1}^2}-1\right).
\end{equation}
From Eqs. (69) and (72) the deviation of $\bar s_W^2$ can be derived 
as
\begin{eqnarray}
\Delta\bar s_W^2&\equiv& \bar s_Z^{\prime 2} - \bar s_W^2 \nonumber \\
&\simeq& -{\bar s^2_W\bar c^2_W\over \bar c^2_W-\bar s^2_W}
\left({m_Z^2\over m_{Z_1}^2}-1\right)+ s_Wc^2_W\xi\tan\chi.
\end{eqnarray}
If we remind the definition of famous $S$ and $T$ parameters \cite{st}
\begin{equation}
\rho =1+\alpha T, \qquad
\Delta\bar s_W^2={\alpha\over c^2_W-s^2_W}\left(-c^2_Ws^2_WT
+{1\over 4}S\right), 
\end{equation}
we can obtain the contributions from the abelian gauge kinetic term
mixing and also the mass mixing to
these parameters as follows,
\begin{eqnarray}
&&\alpha T_M\simeq 2s_W\xi\tan\chi+\left({m_Z^2\over m_{Z_1}^2}-1\right),\\
&&\alpha S_M\simeq 4c^2_Ws_W\xi\tan\chi.
\end{eqnarray} 
Since the difference between $\bar s_W$ and $s_W$ is a higher order
effect in Eqs. (75)-(79), we can replace $s_W$ with $\bar s_W$ in
those formulae for the practical calculation. 


\newpage
\begin{thebibliography}{99}
\bibitem{rev}For a review, see for example, 
H.~-P.~Nilles, Phys. Rep. {\bf C110}, 1 (1984);
H.~E.~Haber and G.~C.~Kane, Phys. Rep. {\bf 117}, 75 (1985).

\bibitem{dflat}For example, M.~Dine, V.~Kaplunovsky, M.~Mangano, C.~Nappi and
N.~Seiberg, Nucl. Phys. {\bf B259}, 549 (1985);
T.~Matsuoka and D.~Suematsu, Nucl. Phys. {\bf B274},
106 (1986); Prog. Theor. Phys. {\bf 76}, 886 (1986).

\bibitem{mu}J.~E.~Kim and H.~P.~Nilles, Phys. Lett. {\bf B138}, 150 (1984).

\bibitem{superb}J.~E.~Kim and H.~P.~Nilles, Phys. Lett. {\bf B263},
79 (1991); 
E.~J.~Chun, J.~E.~Kim and H.~P.~Nilles, Nucl. Phys. {\bf B370}, 105 (1992);
J.~A.~Casas and C.~Mu\~noz, Phys. Lett {\bf B306}, 288 (1993);
G.~F.~Giudice and A.~Masiero, Phys. Lett. {\bf B206}, 480 (1988).

\bibitem{singlet}H.~P.~Nilles, M.~Srednicki and D.~Wyler, Phys. Lett.
{\bf B120}, 346 (1983);
J.-P.~Derendinger and C.~A.~Savoy, Nucl. Phys. {\bf B237}, 307 (1984);
L.~E.~Ib\'a\~nez and J.~Mas, Nucl. Phys. {\bf B286}, 107 (1987);
J.~Ellis, J.~F.~Gunion, H.~E.~Haber, L.~Roszkowski and F.~Zwirner,
Phys. Rev. {\bf D39}, 844 (1989).

\bibitem{rad}K.~Inoue, K.~Kakuto, H.~Komatsu ans S.~Takeshita,
Prog. theor. Phys. {\bf 68}, 927 (1982);
L.~Alvarez-Gaume, J.~Polchinski and M.~B.~Wise, Nucl. Phys. {\bf B221}, 
495 (1983).

\bibitem{extra1}D.~Suematsu and Y.~Yamagishi, Int. J. Mod. Phys. {\bf A10}, 
4521 (1995).

\bibitem{extra2}J.~Hewett and T.~G.~Rizzo, Phys. Rep. {\bf C183}, 193 (1989);
T.~Matsuoka, H.~Mino, D.~Suematsu and S.~Watanabe, 
Prog. Theor. Phys. {\bf 76}, 915 (1986);
F.~Zwirner, Int. J. Mod. Phys. {\bf A3}, 49 (1988).

\bibitem{lang}For a recent review, see for example,
M.~Cveti$\check{\rm c}$ and P.~Langacker, UPR-0761-T, hep-ph/9707451.

\bibitem{cvet}J.~D.~Lykken, Phys. Rev. {\bf D54}, 3693 (1996); 
M.~Cveti$\check{\rm c}$ and P.~Langacker,
Phys. Rev. {\bf D54}, 3570 (1996) and Mod. Phys. Lett. {\bf 11A}, 1247
(1996).

\bibitem{cdeem}M.~Cveti$\check{\rm c}$, D.~A.~Demir, 
J.~R.~Espinosa, L.~Everett 
and P.~Langacker, Phys. Rev. {\bf D56}, 2861 (1997).

\bibitem{wl}P.~Langacker and J.~Wang, hep-ph/9804428.

\bibitem{hold}B.~Holdom, Phys. Lett. {\bf B166}, 196 (1986).

\bibitem{ms}T.~Matsuoka and D.~Suematsu, Prog. Theor. Phys. {\bf 76}, 901
(1986).

\bibitem{mix2}K.~R.~Dienes, C.~Kolda and J.~March-Russell,
Nucl. Phys. {\bf B492}, 104 (1997).

\bibitem{run}F.~del Aguila, G.~D.~Coughlan and M.~Quir\'os,
Nucl. Phys. {\bf B307}, 633 (1988).

\bibitem{electro}B.~Holdom, Phys. Lett. {\bf B259}, 329 (1991).

\bibitem{kinet}K.~S.~Babu, C.~Kolda and J.~March-Russell,
Phys. Rev. {\bf D54}, 4635 (1996); Phys. Rev. {\bf D57}, 6788 (1998).

\bibitem{kinet2}Y.~Umeda, Gi-Chol Cho, K.~Hagiwara, hep-ph/9805447;
Gi-Chol Cho, K.~Hagiwara and Y.~Umeda, hep-ph/9805448.

\bibitem{riz}T.~G.~Rizzo, hep-ph/9806397.
 
\bibitem{neutralino1}D.~Suematsu, Phys. Lett. {\bf B416}, 108 (1998);
Mod. Phys. Lett. {\bf A12}, 1709 (1997).

\bibitem{neutralino2}D.~Suematsu, Phys. Rev. {\bf D57}, 1738 (1998).

\bibitem{mix1}K.~Choi and J.~E.~Kim, Phys. Lett. {\bf 165B}, 71 (1985).

\bibitem{sugra}E.~Cremmer, S~Ferrara, L.~Girardello and A.van Proeyen,
Nucl. Phys. {\bf B212}, 413 (1983).

\bibitem{discr}N.~Haba, C.~Hattori, M.~Matsuda, T.~Matsuoka and
D.~Mochinaga, Prog. Theor. Phys. {\bf 92}, 153 (1994);
G,~Cleaver, M.Cveti$\check{\rm c}$, J.~R.~Espinosa, L.~Everett and
P.~Langacker, Phys. Rev. {\bf D57}, 2701 (1998).

\bibitem{string}I.~Antoniadis, C.~Bachas and C.~Kounnas,
Nucl. Phys. {\bf B289}, 87 (1987); H.~Kawai, D.~Lewellen and
S.~H.~H.~Tye, Phys. Rev. Lett. {\bf 57}, 1832 (1986); Nucl. Phys. {\bf 
B288}, 1 (1987).

\bibitem{sue0}D.~Suematsu, Prog. Theor. Phys. {\bf 96}, 611 (1996).

\bibitem{lang2}P.~Langacker, hep-ph/9805281.

\bibitem{infl}D.~Suematsu and Y.~Yamagishi, Mod. Phys. Lett. {\bf A10},
2923 (1995).

\bibitem{neut}D.~Suematsu, Phys. Lett. {\bf B392}, 413 (1997).

\bibitem{tev}F.~Abe {\it et al.}, CDF Collaboration,
Phys. Rev. Lett. {\bf 79}, 2192 (1997); S.~Abachi {\it et al.}, D0
Collaboration, Phys. Lett. {\bf B385}, 471 (1996).

\bibitem{lep}For example,
M.~Cveti$\check{\rm c}$ and S.~Godfrey, in Proceedings of 
Electro-weak SymmetryBreaking and beyond the standard model, 
eds T.~Barklow, S.~Dawson, H.~Haber and J.~Siegrist 
(World Scientific 1995), hep-ph/9504216;
 K.~Maeshima, Proceedings of the 28th International Conference on High
Energy Physics (ICHEP'96), Warsaw, Poland, 1996;
L3 Collaboration, Phys. Lett. {\bf B306}, 187 (1993);
DELPHI Collaboration, Z. Physik {\bf C65}, 603 (1995).
 
\bibitem{prec}G.~Altarelli, R.~Casalbuoni, D.~Dominici, F.~Feruglio
and R.~Gatto, Mod. Phys. Lett. {\bf A5}, 495 (1990).

\bibitem{nmssm}J.-I.~Kamoshita, Y.~Okada and M.~Tanaka,
Phys. Lett. {\bf B328}, 67 (1994).

\bibitem{particle}C.~Caso {\it et al.}, Particle Data Group, Eur.
Phys. J. {\bf C3}, 1 (1998).

\bibitem{drees}M.~Drees, Phys. Rev. {\bf D35}, 2910 (1987).

\bibitem{st}M.~E.~Peskin and T.~Takeuchi, Phys. Rev. Lett. {\bf 65},
964 (1990); Phys. Rev. {\bf D46}, 381 (1992).  

\end{thebibliography}
\end{document}